\documentclass[prb,twocolumn,superscriptaddress,showpacs,nofootinbib]{revtex4-1}
\usepackage{amsmath,amssymb,amsfonts,amsthm}
\usepackage{graphicx}
\usepackage{bm}
\usepackage{bbm}
\usepackage{color}
\usepackage[usenames,dvipsnames]{xcolor}
\usepackage{dcolumn}   
\usepackage{epstopdf}
\usepackage[caption=false]{subfig}
\usepackage{setspace}
\usepackage{soul}
\usepackage{color}
\usepackage[colorlinks=true,linkcolor=blue,urlcolor=blue,citecolor=blue]{hyperref}
\usepackage{bbold}

\begin{document}

 \newcommand{\breite}{1.0} 

\newtheorem{prop}{Proposition}
\newtheorem{cor}{Corollary}

\newcommand{\be}{\begin{equation}}
\newcommand{\ee}{\end{equation}}

\newcommand{\bea}{\begin{eqnarray}}
\newcommand{\eea}{\end{eqnarray}}
\newcommand{\lt}{<}
\newcommand{\gt}{>}

\newcommand{\Reals}{\mathbb{R}}     
\newcommand{\Com}{\mathbb{C}}       
\newcommand{\Nat}{\mathbb{N}}       

\newcommand{\mkch}[1]{{\color{BrickRed} #1}}

\newcommand{\id}{\mathbboldsymbol{1}}

\newcommand{\Real}{\mathop{\mathrm{Re}}}
\newcommand{\Imag}{\mathop{\mathrm{Im}}}

\def\O{\mbox{$\mathcal{O}$}}   
\def\F{\mathcal{F}}			
\def\sgn{\text{sgn}}

\newcommand{\deo}{\ensuremath{\Delta_0}}
\newcommand{\dea}{\ensuremath{\Delta}}
\newcommand{\ak}{\ensuremath{a_k}}
\newcommand{\ad}{\ensuremath{a^{\dagger}_{-k}}}
\newcommand{\sx}{\ensuremath{\sigma_x}}
\newcommand{\sz}{\ensuremath{\sigma_z}}
\newcommand{\spl}{\ensuremath{\sigma_{+}}}
\newcommand{\smi}{\ensuremath{\sigma_{-}}}
\newcommand{\alk}{\ensuremath{\alpha_{k}}}
\newcommand{\bk}{\ensuremath{\beta_{k}}}
\newcommand{\ok}{\ensuremath{\omega_{k}}}
\newcommand{\vd}{\ensuremath{V^{\dagger}_1}}
\newcommand{\vi}{\ensuremath{V_1}}
\newcommand{\vo}{\ensuremath{V_o}}
\newcommand{\zc}{\ensuremath{\frac{E_z}{E}}}
\newcommand{\xc}{\ensuremath{\frac{\Delta}{E}}}
\newcommand{\xd}{\ensuremath{X^{\dagger}}}
\newcommand{\aok}{\ensuremath{\frac{\alk}{\ok}}}
\newcommand{\tpw}{\ensuremath{e^{i \ok s }}}
\newcommand{\tpe}{\ensuremath{e^{2iE s }}}
\newcommand{\tmw}{\ensuremath{e^{-i \ok s }}}
\newcommand{\tme}{\ensuremath{e^{-2iE s }}}
\newcommand{\epls}{\ensuremath{e^{F(s)}}}
\newcommand{\emis}{\ensuremath{e^{-F(s)}}}
\newcommand{\epl}{\ensuremath{e^{F(0)}}}
\newcommand{\emi}{\ensuremath{e^{F(0)}}}

\newcommand{\lr}[1]{\left( #1 \right)}
\newcommand{\lrs}[1]{\left( #1 \right)^2}
\newcommand{\lrb}[1]{\left< #1\right>}
\newcommand{\nbt}{\ensuremath{\lr{ \lr{n_k + 1} \tmw + n_k \tpw  }}}

\def\beq{\begin{equation}}
\def\eeq{\end{equation}}
\def\bea{\begin{eqnarray}}
\def\eea{\end{eqnarray}}

\newcommand{\om}{\ensuremath{\omega}}
\newcommand{\dw}{\ensuremath{\Delta_0}}
\newcommand{\wbp}{\ensuremath{\omega_0}}
\newcommand{\dv}{\ensuremath{\Delta_0}}
\newcommand{\vbp}{\ensuremath{\nu_0}}
\newcommand{\vplus}{\ensuremath{\nu_{+}}}
\newcommand{\vminus}{\ensuremath{\nu_{-}}}
\newcommand{\wplus}{\ensuremath{\omega_{+}}}
\newcommand{\wminus}{\ensuremath{\omega_{-}}}
\newcommand{\uv}[1]{\ensuremath{\mathbf{\hat{#1}}}} 
\newcommand{\abs}[1]{\left| #1 \right|} 
\newcommand{\avg}[1]{\left< #1 \right>} 
\let\underdot=\d 
\renewcommand{\d}[2]{\frac{d #1}{d #2}} 
\newcommand{\dd}[2]{\frac{d^2 #1}{d #2^2}} 
\newcommand{\pd}[2]{\frac{\partial #1}{\partial #2}}
\newcommand{\pdd}[2]{\frac{\partial^2 #1}{\partial #2^2}}
\newcommand{\pdc}[3]{\left( \frac{\partial #1}{\partial #2}
 \right)_{#3}} 
\newcommand{\ket}[1]{\left| #1 \right>} 
\newcommand{\bra}[1]{\left< #1 \right|} 
\newcommand{\braket}[2]{\left< #1 \vphantom{#2} \right|
 \left. #2 \vphantom{#1} \right>} 
\newcommand{\matrixel}[3]{\left< #1 \vphantom{#2#3} \right|
 #2 \left| #3 \vphantom{#1#2} \right>} 
\newcommand{\grad}[1]{{\nabla} {#1}} 
\let\divsymb=\div 
\renewcommand{\div}[1]{{\nabla} \cdot \boldsymbol{#1}} 
\newcommand{\curl}[1]{{\nabla} \times \boldsymbol{#1}} 
\newcommand{\laplace}[1]{\nabla^2 \boldsymbol{#1}}
\newcommand{\vs}[1]{\boldsymbol{#1}}
\let\baraccent=\= 

\newcommand{\sg}[1]{{\color{red} #1}}
\newcommand{\dah}[1]{{\color{blue} #1}}
\newcommand{\mk}[1]{{\color{cyan} #1}}
\newcommand{\ka}[1]{{\color{magenta} #1}}
\newcommand{\ea}[1]{{\color{olive} #1}}


\title{Localization and transport in a strongly driven Anderson insulator}

\author{Kartiek Agarwal}
\affiliation{Department of Electrical Engineering, Princeton University, Princeton NJ 08544, USA}%
\author{Sriram Ganeshan}
\affiliation{Simons Center for Geometry and Physics, Stony Brook, NY 11794, USA}%
\author{R. N. Bhatt}
\affiliation{Department of Electrical Engineering, Princeton University, Princeton NJ 08544, USA}%

\date{\today}
\begin{abstract}

We study localization and charge dynamics in a monochromatically driven one-dimensional Anderson insulator focussing on the low-frequency, strong-driving regime. We study this problem using a mapping of the Floquet Hamiltonian to a hopping problem with correlated disorder in one higher harmonic-space dimension. We show that (i) resonances in this model correspond to \emph{adiabatic} Landau-Zener (LZ) transitions that occur due to level crossings between lattice sites over the course of dynamics; (ii) the proliferation of these resonances leads to dynamics that \emph{appear} diffusive over a single drive cycle, but the system always remains localized; (iii) actual charge transport occurs over many drive cycles due to slow dephasing between these LZ orbits and is logarithmic-in-time, with a crucial role being played by far-off Mott-like resonances; and (iv) applying a spatially-varying random phase to the drive tends to decrease localization, suggestive of weak-localization physics. We derive the conditions for the strong driving regime, determining the parametric dependencies of the size of Floquet eigenstates, and time-scales associated with the dynamics, and corroborate the findings using both numerical scaling collapses and analytical arguments. 

\end{abstract}
\maketitle

\section{Introduction}The advent of new technologies to control and probe artificial quantum matter in extremely isolated settings has generated immense interest in the study of non-equilibrium dynamics of quantum systems~\cite{Greiner,Kinoshita06,Langen,schreiber2015observation,smith2016many}. Much of this interest has been directed at the study of quantum quench protocols~\cite{Kinoshita06,Cheneau,Langen,sadler2006spontaneous}. On the other hand, coherent periodic driving---naturally achieved using lasers in experimental setups---has been primarily used to adiabatically engineer novel Hamiltonians~\cite{cooperfluxlattice,kitagawa2011transport,galitski2013spin,dalibardguagepotentials,kennedy2015observation}. The reason for this is that in the former case the energy pumped into the system is bounded---one can then study associated questions of re-equilibration, dynamical instabilities, novel correlations, etc.---while for the latter it was until recently believed~\cite{LucaIsolatedDriven,dalessio2013heating,ponte2015periodically,rehnheatingmbl} that periodically driven systems eventually heat up to infinite temperature, forming an incoherent soup devoid of novel properties.  

Recent work has shown that periodically driven systems can avoid heating up indefinitely, or take an inordinately long time to do so, and exhibit interesting physics entirely novel to them. These include the discovery of long-lived prethermal states in driven quantum systems~\cite{prethermallonglived,abanin2015effective,abanin2015rigorous,bukovtwoband}, the Floquet many-body-localized (MBL) phase~\cite{pontefloquetMBL,lazaridesperiodicmbl,abanin2016theory}, discrete time crystals~\cite{vedikatimecrystal,elsetimecrystal,keyserlingktimecrystalstability,YaoRigidity}, re-entrant MBL phases that emerge only due to driving~\cite{bairey2017driving,ho2017critical}, integrable Floquet phases~\cite{gritsev2017integrable}, and new topological phases~\cite{martin2016topological,potter2016topological,Pochiralfloquetbosons,TitumFloquetAnderson,nathan2016quantized}. Further, experiments (including many notable ones in traditional condensed matter systems, see Refs.~\cite{fausti2011light,wang2013observation,levonian2016probing}) have both corroborated some of these findings~\cite{zhang2017observation,choi2017observation,bordia2016periodically,floquetprethermalregime} and provided examples~\cite{houck2012chip,schecter2012dynamics,meinert2016bloch,baumann2010dicke} of new dynamical phenomena in strongly driven settings. Thus, periodic driving has emerged as a new tool in understanding the non-equilibrium properties of quantum matter. 

While much progress has been made in establishing the existence of localized (that naturally escape the heat death) Floquet phases, much less is understood about the precise nature of localization and dynamics in such systems. Semi-analytic arguments exist that outline the conditions wherein the Floquet-MBL phase should get destabilized by a strong periodic drive~\cite{abanin2016theory} or display non-linear heating/dynamical regimes~\cite{heatingmbldrivenSG}, but a more quantitative, and detailed analysis of the strong-driving regime has not been carried out. A recent work~\cite{ducatez2016anderson} provides conditions on the stability of a weakly driven Anderson insulator, but does not delve into the strong-driving regime. 

In this work we treat the problem of periodically driving a disordered system on the simpler platform of a one-dimensional Anderson insulator. Using methods first developed by J. Shirley~\cite{shirleyTLS}, this problem can be mapped on to a quasi-one-dimensional single-particle hopping problem, in an additional harmonic-space dimension [besides the real space dimension(s), see Fig.~\ref{fig:phasefig}] which is effectively limited to a width $\sim 1/\omega$. This provides both a better analytical handle of the problem and allows for accurate numerics to corroborate analytical expectations. It is also suitable for treating the strong driving regime accurately which stands in contrast to the high-frequency Magnus expansion~\cite{blanes2009magnus}. 

 In the weak-driving regime, wherein the coupling to the drive is weak, or the drive-frequency is high, the system remains localized on length scales of the static localization length. As we show, in the higher-dimensional description of the problem, this regime corresponds to the situation where only isolated resonances exist between different lattice sites. The strong-driving regime is obtained when these resonances proliferate. However, the proliferation of these resonances does not lead to delocalization in the one-dimensional Anderson model---this is clear from the quasi-one-dimensional character of the Floquet-Hamiltonian description of the problem, and the prevalence of localization in one dimension for arbitrarily weak disorder~\cite{absenceofdiffusionanderson,scalingtheoryandersonlocalization}. Understanding the fate of localization and dynamics in this regime is the main goal of this work. 

First, we show that resonances in this higher-dimensional model correspond to \emph{adiabatic} Landau-Zener (LZ) transitions---as the local potential is varied over a drive cycle, level crossings between pairs of sites occur and only those pairs among these which involve adiabatic charge movement correspond to the resonances. The typical length scale at which these resonances occur is $x_{\text{ad}} = \frac{\xi}{2} \text{log} (W^2/A \omega)$, where $\xi$ is the static localization length, $W$ is the typical disorder strength (of the order of the tunneling amplitude), $A$ is the drive amplitude and $\omega$ is the drive frequency. As mentioned above, the localization properties of the system are not altered until such resonances proliferate; this occurs above a critical amplitude $A_c$ for which $x_{\text{ad}} \lesssim x_{\text{LZ}}$, where $x_{\text{LZ}} \approx W/A$ is the length at which one certainly finds a level crossing between sites in the disordered landscape of the Anderson model. We find clear numerical evidence showing that this critical drive amplitude depends only logarithmically on the drive frequency, and linearly on the tunneling strength, as suggested by these expressions. 

\begin{center}
\begin{figure}
\includegraphics[width = 0.47 \textwidth]{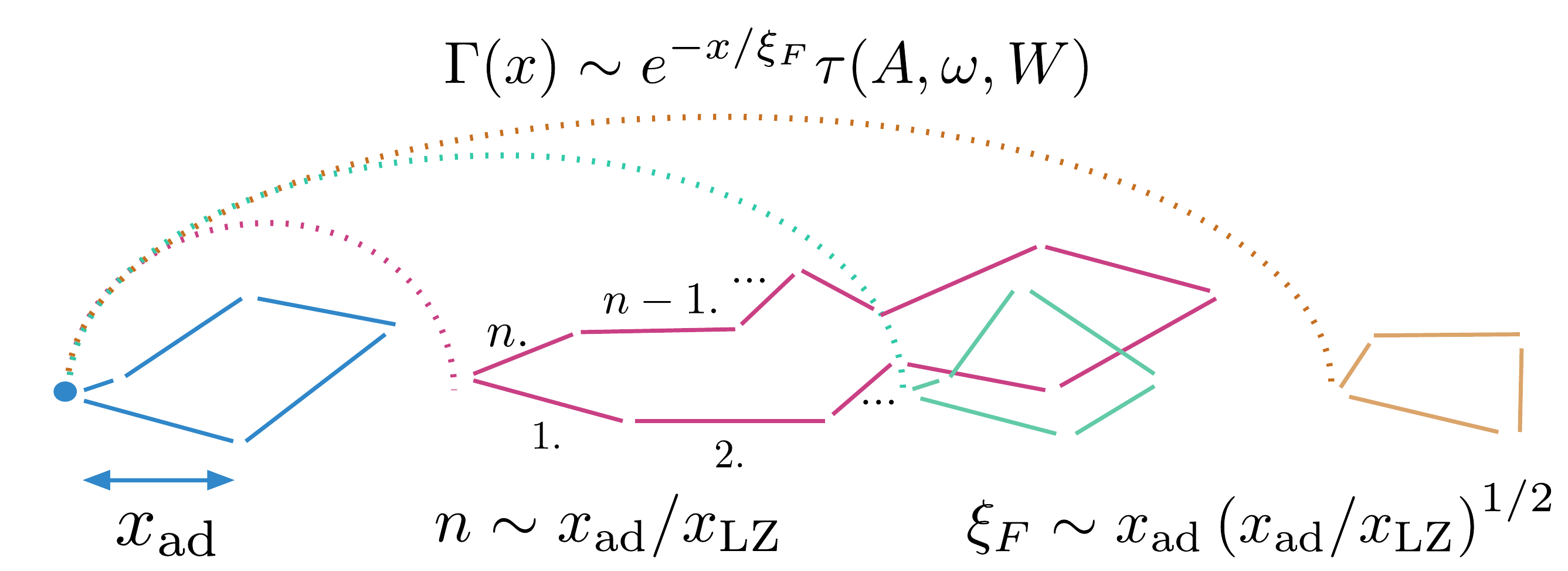} 
\caption{A picture illustrating the physics of the strong driving regime. Landau-Zener transitions proliferate, but the Floquet eigenstates are still localized and form from closed circuits of these LZ transitions. The size of these orbits is given by the diffusive spread of charge on the timescale of a single drive cycle. Actual charge transport over many drive cycles occurs due to dephasing between Mott-like pairs of these orbits. This occurs at time-scales exponentially large in the distance moved.}
\label{fig:mottdance}
\end{figure}
\end{center}

In the strong driving regime, we argue that Floquet eigenstates are localized on a length $\xi_F \sim x_{\text{ad}} \left( x_{\text{ad}} / x_{\text{LZ}} \right)^{1/2}$; this comes from the picture that Floquet eigenstates are closed LZ `circuits', wherein the charge performs random walk in the space of LZ transitions, eventually returning to the origin after a complete drive cycle. Since the expected number of LZ in each cycle is given by $x_{\text{ad}} / x_{\text{LZ}}$, and each step is of the length $x_{\text{ad}}$, the distance `diffused' in one drive cycle gives the size of the Floquet eigenstates. We find excellent scaling collapse of the numerical data (see Fig.~\ref{fig:xifscaling}) that corroborates this picture. We provide a caricature of this physics in Fig.~\ref{fig:mottdance}.

\begin{center}
\begin{figure*}
\includegraphics[width=1 \textwidth]{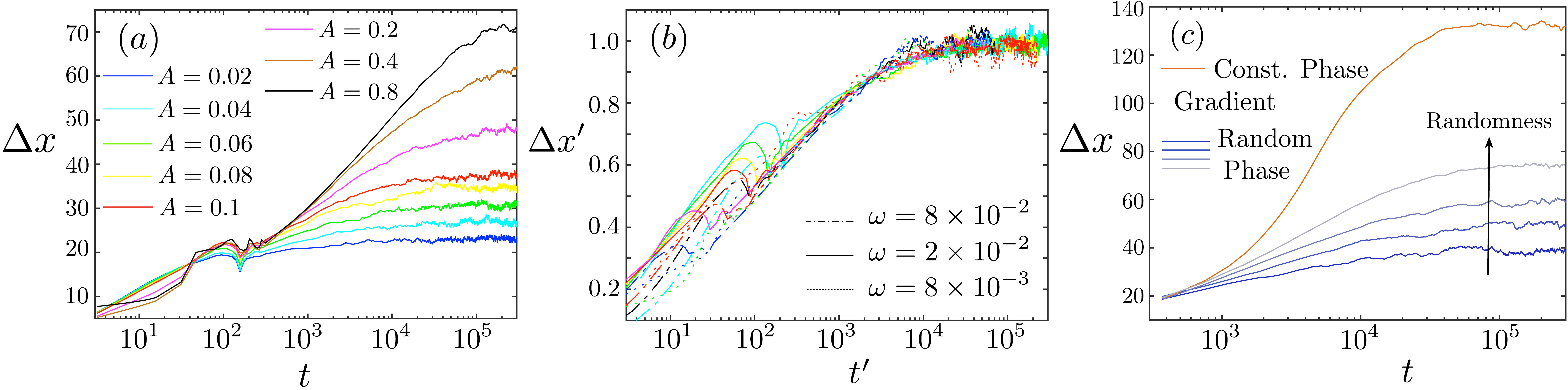}
\caption{Dynamics in a driven Anderson Insulator. (a) Logarithmic growth of \emph{rms} distance $\Delta x$ in time $t$ shown for the case $\omega = 0.02$, $W = 1$. The logarithmic occurs beyond the single drive cycle time, $T_\omega = 100 \pi$ in this case. (b) Scaling collapse of logarithmic-in-time dynamics by rescaling length and time axes: $\Delta x' = \Delta x /\xi_F$ and $ t' = t \sqrt{W^3 T_\omega} / A$. The collapse is performed for $W =1$, and multiple drive frequencies and amplitudes (corresponding to the strong driving regime). Note that the departure from the logarithmic group at small times $t'$ is associated with the real-time dynamics for times $t \lesssim T_\omega$. (c) Weak localization of LZ orbits: application of a random spatially-varying drive phase results in larger LZ orbits. The application of a non-random phase with a fixed spatial gradient develops topological properties and shows diffusive dynamics instead (while still being localized on a much larger scale not capture in this finite-size simulation).}
\label{fig:figdyn}
\end{figure*}
\end{center}

A further important observation is the logarithmic spread of charge over many drive cycles, measured stroboscopically [Fig.~\ref{fig:figdyn} (a)]. As we explain below, this logarithmic spreading likely occurs due to dephasing between Mott-like pairs~\cite{mottconductivity} of various LZ orbits (equivalently, Floquet eigenstates) at a rate $\Gamma(x) \sim e^{-x/\xi_F} \tau( A, \omega, W)$, where the exponential dependence in $x$ comes from the fact that the matrix element for a local position operator connecting different Floquet eigenstates distance $x$ apart decays exponentially (in $x$). Note also that the logarithmic-in-time growth emerges most clearly after averaging over many initial starting locations of the charge. Moreover, even though the charge displacement saturates at long times at a distance of the order of $\xi_F$, the full probability distribution of the charge displacement [see Fig.~\ref{fig:dynextra} (a)] has considerable weight at farther distances. This again shows the influence of `rare' long-distance Mott pairs in the dynamics. The timescale $\tau(A,\omega,W)$ can be expressed in terms of $\xi_F$, and suitable Bessel functions; it scales as $\tau(A,\omega,W) \sim \sqrt{\frac{A}{W^{3/2}\omega^{1/2}}}$, and we verify this dependence via a scaling collapse [Fig.~\ref{fig:figdyn} (b)] of the data. 

We note that the picture of LZ orbits within the confines of which the particle performs diffusive motion, but is ultimately localized is reminiscent of weak-localization physics. This picture is reinforced by the observation that an application of the drive with a random,  spatially-varying phase enhances the size of these orbits without changing the essential physics [Fig.~\ref{fig:figdyn} (c)]. We conclude by a limited discussion of some aspects of the problem in higher dimensions, and multi-harmonic drives. 

\section{Model And Observables}

 We study the one-dimensional Anderson model, $H_{\text{Anderson}} = \sum_x \epsilon_x f^\dagger_x f_x - t f^\dagger_{x+1} f_x - h.c.$ with local energies $\epsilon_x$ drawn uniformly $ \in [-W/2,W/2]$, under the application of a periodic drive that modulates the local potential in a staggered manner: $H_{\text{Drive}} = 2 \sum_x f^\dagger_x f_x A \text{cos} \left\{\omega t + [ \pi + \delta(x)] x \right\}$. In what follows, we set $t = 1$. Note that $\delta(x) = \text{const.} \neq 0$ is special as it lends a chirality to the drive which can dramatically change the properties of the system for large $\delta$; this case will be discussed later. We assume for now that $\delta(x) = 0$ although our results apply equally to the case when $\delta(x) \ll 1$ and random. 

\begin{center}
\begin{figure}
\includegraphics[width=0.46 \textwidth]{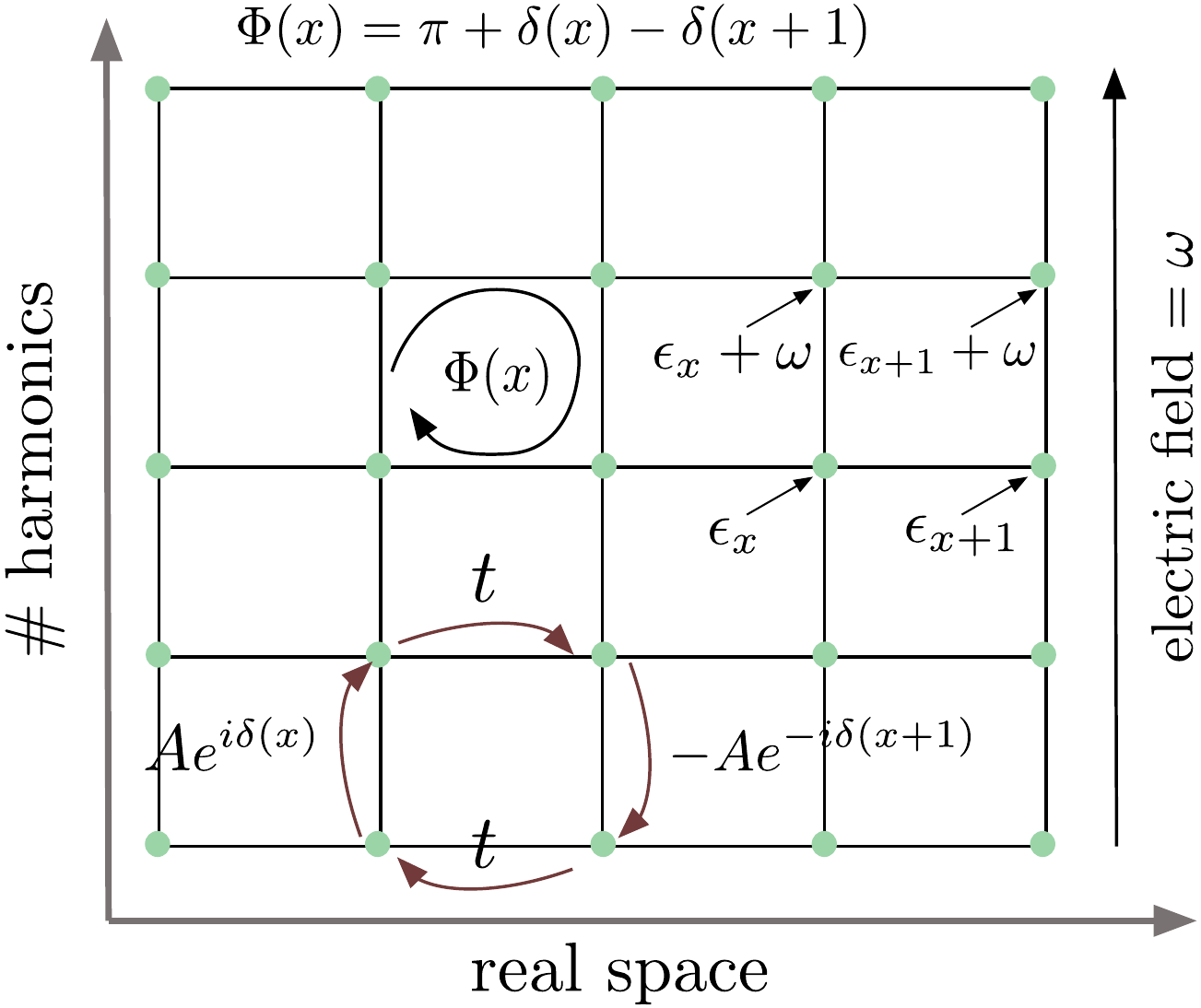}
\caption{The Floquet Hamiltonian. Hopping in the real space direction occurs with the real parameter $t = 1$, while hopping in the harmonic space direction occurs with an amplitude proportional to the drive strength $A$, and a phase $\pi x + \delta(x)$. A drive with phase $\delta(x) \neq 0$ shows up as an effective magnetic field in the Floquet Hamiltonian.}
\label{fig:phasefig}
\end{figure}
\end{center}

The Floquet eigenfunctions are given by Bloch's theorem by $\ket{\psi_\alpha (t)} = e^{-i \epsilon_\alpha t} \ket{\phi_\alpha (t) } $, where $\ket{\phi_\alpha(t)}$ is a periodic function of time, with period $T_\omega = 2\pi/\omega$, and $\epsilon_\alpha$ is the quasi-energy defined modulo $\omega$. $\ket{\phi_\alpha(t)}$ can be expanded in the basis of harmonics of $\omega$, and real-space coordinates, denoted by $\ket{n,x}$: $\ket{\phi_\alpha(t)} = \sum_{n,x} \phi_\alpha(n,x) \ket{n,x} e^{i n \omega t}$ where the coefficients $\phi_\alpha (n,x)$ are given by eigen-solutions of the following Floquet Hamiltonian $H_F$ (for a derivation see for example Refs.~\cite{shirleyTLS,ducatez2016anderson}): 

\begin{align}
&H_F = \sum_{n,x} A [ e^{i \pi + i \delta(x)} c^\dagger_{n,x} c_{n+1,x} + h.c.], \nonumber \\
& - t \sum_x [c^\dagger_{n,x} c_{n,x+1} + h.c.] + \sum_{n,x} (\epsilon_x + n \omega) c^\dagger_{n,x} c_{n,x}, 
\label{eq:defhf}
\end{align}

where we will set $t = 1$. $H_F$ has an extremely simple physical interpretation in terms of the number $n$ of photons of the drive: each photon costs as energy $\omega$, and consequently, the local potential at any site $(n,x)$ is $\epsilon_x + n \omega$. The drive $A_0 e^{i \omega t} + A^{*}_{0} e^{-\omega t}$ doesn't change the position (since it couples to the local density) of the particle but can give or absorb a harmonic, with amplitude $A_0$ or $A^*_0$, and shows up in terms such as $c^\dagger_{n,x} c_{n+1,x}$. The operators $c^\dagger_{n,x}$ and $c_{n,x}$ retain the statistics of the original particles in the Anderson system since the absorption of any number of bosons does not change this. A drive with a phase $\delta(x)$ shows up as an effective magnetic field in the problem. These features are illustrated in Fig.~\ref{fig:phasefig}. 

The Hamiltonian has $L$ (real-space dimension) unique eigenfunctions which can be found by exact diagonalization. All other eigenstates are constructed by translation in harmonic space, and a corresponding increase in energy by appropriate multiples of $\omega$.  Thus, we diagonalize $H_F$ restricting the number of harmonics to $n \in [-N/2,N/2]$, where $N \sim W/\omega$. 

\begin{center}
\begin{figure*}
\includegraphics[width=1 \textwidth]{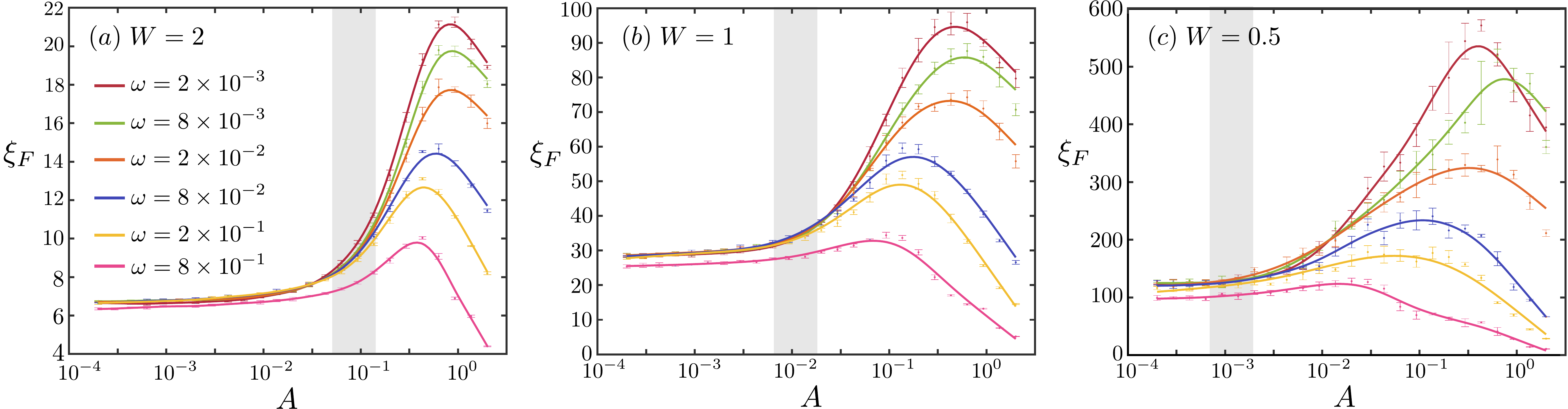}
\caption{Floquet localization lengths $\xi_F$ calculated using the transfer matrix method for different drive settings (frequency $\omega$, amplitude $A$) and disorder strengths (a) $W =2$, (b) $W = 1$, and (c) $W = 0.5$. The shaded regions describe the critical drive amplitude at which the system should occur the strong driving regime, as per the picture of proliferation of resonances in the Floquet Hamiltonian.}
\label{fig:xif}
\end{figure*}
\end{center}

\begin{center}
\begin{figure}
\includegraphics[width=0.46 \textwidth]{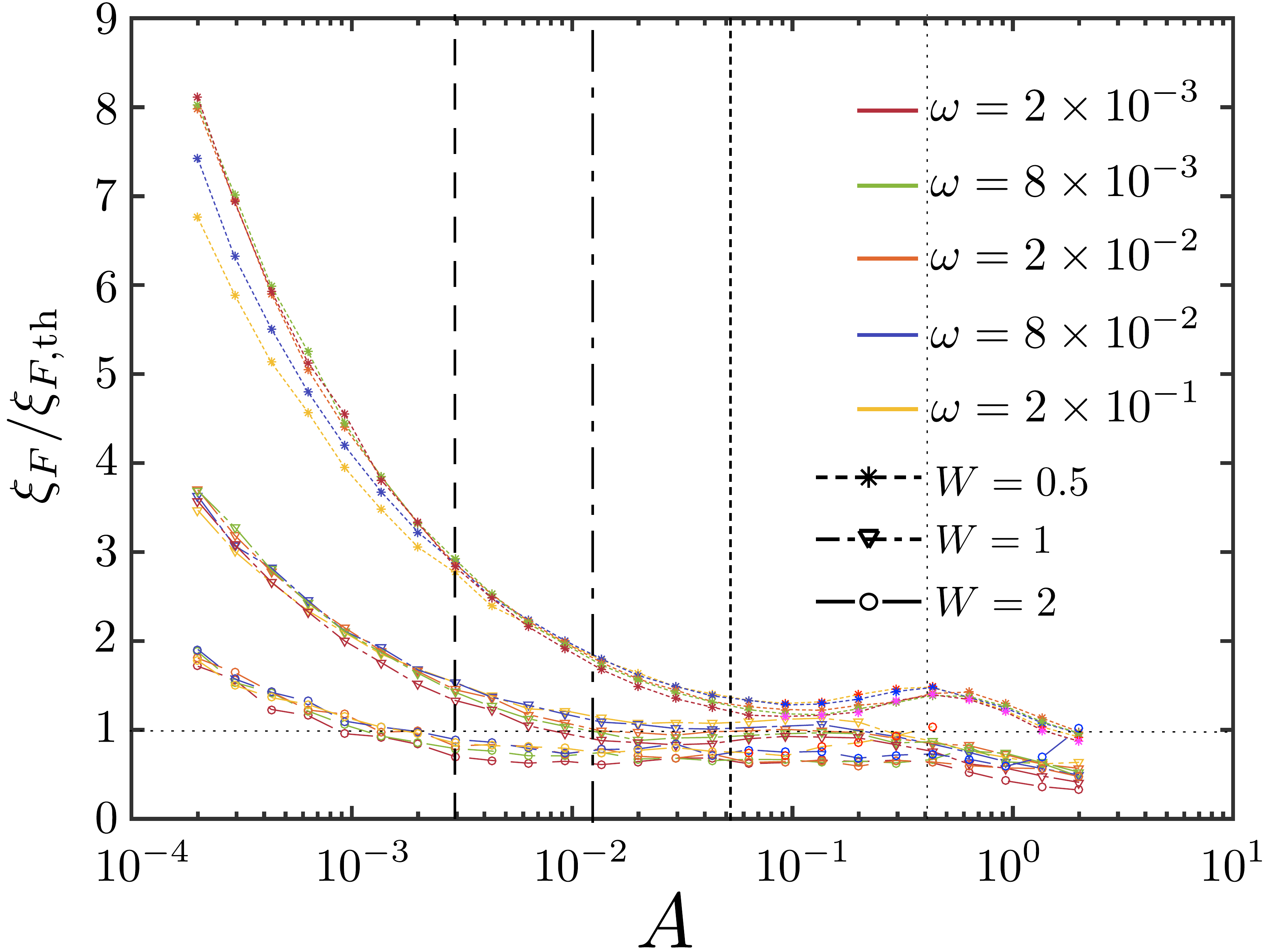}
\caption{Scaling collapse of the Floquet localization length $\xi_F$ shown in Fig.~\ref{fig:xif} according to the scaling $\xi_{F,\text{th}} = c(\omega, W) x_{\text{ad}} \left( x_{\text{ad}} / x_{\text{LZ}} \right)^{1/2}$. $c(\omega, W)$ is determined by the conditions $x_{\text{ad}} (A_c) = x_{\text{LZ}} (A_c)$, and $\xi_{F,\text{th}} (A_c) = \xi$, where $\xi$ is the static localization length. The vertical dashed lines delineate (approximately, ignoring logarithmic frequency dependence) the critical amplitude $A = A_c$ above which the strong driving regime holds for the different disorder strengths $W = 0.5,1,2$.}
\label{fig:xifscaling}
\end{figure}
\end{center}

To compute the Floquet localization length $\xi_F$, we use the transfer matrix method~\cite{mackinnonkramer,leetwoDanderson}. In this case, the system is truly quasi-one-dimensional and so the transfer matrix method can be used directly without resorting to scaling assumptions. 
For completeness, we note that transfer-matrix approach calculates the transmittance between the left end (in real space) of this quasi-1D strip to the right end using a recursive approach that in particular, allows one to take $L$ as large as one wishes (in practice, until desired accuracy in computing $\xi_F$ is achieved); see Ref.~\cite{mackinnonkramer} for details. In terms of the resolvent matrix $G(L) = (E - H)^{-1}$ connecting the $N$ harmonic-space sites at the left end of the sample to those at the right end, the Floquet localization length is given by $\xi^{-1}_F = \lim_{L \rightarrow \infty} [2 (L-1)]^{-1} \text{ln Tr} \left[ G G^\dagger \right]$. Finally, we note the following subtle point. The use of this method for the dynamical problem we consider here is predicated on the basis that the probability of finding a particle at some site $x$ depends only the magnitudes $\abs{\phi_\alpha (n,x)}^2$ (for all harmonics $n$). This turns out to be true in a time-averaged sense: average occupation of a particle at site $x$ over a drive cycle is given by $\abs{\bar{\phi}_\alpha(x)}^2 = \frac{1}{T} \int_0^T \abs{\braket{x}{\psi_\alpha}}^2 = \sum_{n} \abs{\phi_\alpha (n,x)}^2$. As we will show, the phase information between these coefficients is crucial to obtaining the dynamics.

One can compute the dynamics of charge as follows. We denote by $\ket{\psi_\alpha}$ as one of the $L$ unique eigenfunctions of $H_F$, and $\ket{\psi_{\alpha + m\omega}}$ as the translation of the this wave-function by $m$ harmonics. The completeness relations read $ \braket{\psi_{\alpha + m \omega}}{\psi_{\beta + m' \omega}}  = \delta_{mm'} \delta_{\alpha \beta}$. Assuming the system starts in an initial state $\ket{x_0, n = 0}$, the wave-function at time $t = 0$, is given by $\ket{\psi_{x_0}} (t = 0) = \sum_{m,\alpha} \braket{\psi_{\alpha - m \omega}}{x_0, 0}  = \sum_{\alpha, m} \phi^*_\alpha (x_0, m) \ket{\psi_{\alpha - m \omega}}$. At time $t$, $\ket{\psi_{x_0} (t)} = \sum_{\alpha, m} \phi^*_\alpha (x_0, m) e^{i m \omega t - i \epsilon_\alpha t} \ket{\psi_{\alpha - m \omega}}$. Now, the probability $P(x_t | x_0)$ to find the particle at site $x_t$ at time $t$ is given by $P(x_t | x_0)  = \abs{\sum_{n} \matrixel{x_t, -n}{e^{-i n \omega t}}{\psi_{x_0} (t) } }^2$. This finally yields

\begin{align}
P(x_t | x_0) &= \abs{\sum_\alpha e^{-i \epsilon_\alpha t} O^*_{\alpha} (x_0,t) O_\alpha (x_t,t) }^2, \nonumber \\
O_\alpha (x,t) &= \sum_n \phi_\alpha (x,n)e^{-i n \omega t}.
\end{align}
For stroboscopic measurements, $t = l T$, the expression is simplified further as all internal exponents $e^{\pm i n \omega t}$ evaluate to $1$. 
Using completeness relations one can show that $\sum_{x_t} P(x_t |x_0) = 1$ for all times. Combined with that fact that $P(x_t|x_0) > 0$ for all $x_t$, shows that $P(x_t|x_0)$ is indeed a probability distribution, as expected. From this distribution, one can surmise the rms distance $\Delta x (t) $ travel by the charge in time $t$, as $\sqrt{\sum_{x_t} (x_t -x_0)^2 P(x_t | x_0) }$. We further note that, reassuringly, the precise choice of harmonic configuration of the initial state [in this case, occupation of only $n = 0$ harmonic in $\ket{\psi_{x_0} (t = 0)}$] \emph{does not} alter the result for this probability distribution. 

\section{Numerical Results} 

We first present the results for the Floquet localization length. The calculations have been performed using the transfer-matrix approach, using lengths $L \sim 80,000$ and a harmonic-space width of about $200-800$ harmonics, with increasingly larger widths used until convergence is obtained. Fig.~\ref{fig:xif} shows the Floquet localization length for a number of drive amplitudes $A \in [10^{-4}, 1] W$ and frequencies $\omega \in [10^{-3} ,1] W$, and different disorder strengths $W = 0.5,1,2$. We see that the Floquet localization length does not deviate significantly from the static localization length below a critical drive amplitude $A_c$. $A_c$ is seen to depend weakly on the drive frequency, but varies rapidly with the disorder strength $W$. In the strong driving regime, that is, for $A > A_c$, the Floquet localization length grows rapidly with increasing drive amplitude, but the growth itself depends weakly on the drive frequency. At drive amplitudes $A \sim W$, the Floquet localization length decreases again to values of the order of the static localization length (or lower). This likely occurs in a manner similar to that discussed in Refs.~\cite{bairey2017driving,ho2017critical} where tunneling is strongly suppressed by the drive; we do not investigate this regime in detail and instead focus on the regime of strong driving where both drive amplitude and frequency are smaller than the single-particle bandwidth $\sim W$. In this regime, a scaling collapse can be obtained for the Floquet localization length, as show in Fig.~\ref{fig:xifscaling}. The dominant behavior of $\xi_F \sim \sqrt{A}$, $\text{ln} (\omega)$, will be explained in detail later. 

We next discuss the time-dependent charge spread. Fig.~\ref{fig:figdyn} (a) shows the rms distance traveled by the charge as a function of time for some drive parameters. These results have been obtained by exact diagonalization of the Floquet Hamiltonian, for system sizes $L \le 1200$, and harmonic-space width $N \approx 4 W/\omega \le 800$; convergence occurs for $L \gg \xi_F$ indicating the role of long-distance Floquet eigenstates in charge spreading. The Hilbert space dimension, $L \times N$, of the matrix encoding $H_F$ can thus reach $\sim 10^6$ in our simulations, which is prohibitively large for performing exact diagonalization. We use built-in sparce-matrix routines in MATLAB to extract the unique $L$ eigenfunctions in the energy interval $\in [-\omega/2,\omega/2]$. Convergence is obtained for $N = N_0$ such that for any $N > N_0$, the algorithm finds precisely $L$ eigenfunctions in the above energy interval.   

Note that the averaging procedure is important for obtained a pure logarithmic growth and has some analogy to rare region physics in quenched disorder settings~\cite{agarwalmblprl,agarwal2017rare}. We perform averaging over all initial starting sites for a given disorder realization and over $10$ disorder realizations. An initial rapid spread up to a distance $\sim \xi$ occurs ballistically within a single drive cycle, followed by a slow logarithmic-in-time spread over many hundreds of drive cycles, ultimately saturating at a distance of $\sim \xi_F$. Fig.~\ref{fig:figdyn} (b) shows the scaling collapse of slow logarithmic spread in terms of the following variables: $x' = x/x_{\text{rms}} (t = \infty)$ and $t' = t/\tau$, where $\tau = \frac{A}{\sqrt{\omega} W^{3/2}}$. Note that only data from $W = 1$ was used in the scaling collapse. Fig.~\ref{fig:dynextra} (a) shows the full probability distribution $P(x_t | x_0 = 0)$ for various times (exponentially spaced). The distribution is always peaked at the origin, indicating localization, while the exponential envelope grows broader exponentially slowly in time. Note that the probability distribution is non-zero at scales $\gg \xi_F$, which illustrates the importance of faraway resonances in the spread of charge. 

We also investigate changes in the dynamics due to a drive with random spatially-varying phase, the results of which are shown in Fig.~\ref{fig:figdyn} (c); as the randomness is increased, charge-spread continues be logarithmically in time, but the infinite-time displacement is greater. Thus, we see that phase coherence in the driven plays an important role in determining the size of the Floquet eigenstates, which is suggestive of weak-localization behavior. In Fig.~\ref{fig:figdyn} (c), we also provide an example of the case when the phase of the drive varies linearly in space, that is $\delta (x) \sim q x, q \neq 0, \pi$. This case is markedly differently from the random phase case, as charge spread now occurs diffusively ($\Delta x \sim \sqrt{t}$). As apparent from Fig.~\ref{fig:phasefig}, this case corresponds to the situation where the Floquet Hamiltonian has a constant magnetic field. Given the finite size of the system in the harmonic-space direction, the system develops edge states. The charge can scatter from one edge to another due to disorder, and hence spreads diffusively. 

\section{Resonances in the Floquet Hamiltonian}

 We now explain the above findings by first analyzing the resonance structure in the Floquet Hamiltonian, and subsequently providing the physical picture of these resonances in terms of adiabatic LZ transitions and dephasing between various spatially separated Floquet eigenstates. 

To begin with, we consider the toy model 
\be
H = - \frac{\Delta}{2} \sigma_x - \frac{\epsilon_0}{2} \sigma_z - \frac{A}{2} \sigma_z \left(e^{i \omega t} + e^{-i \omega t} \right).
\ee
 The Pauli-spin operators operate on the two real-space sites in this toy model. If $\Delta = 0$, the eigenfunctions $\xi_{L,k}$ and $\xi_{R,k'}$ for the left $L$ and right $R$ sites, are given by~\cite{shevchenko2010landau,sonLZsolution} 
 \begin{align}
 \xi_{L,k} = \begin{pmatrix} . \\ . \\ . \\ J_{2} (A/\omega) \\ J_1 (A/\omega) \\ J_0 (A/\omega) \\ J_1 (A/\omega) \\ J_2 (A/\omega) \\ . \\ . \end{pmatrix} \; \; \;  \xi_{R,k'} = \begin{pmatrix} . \\ . \\ . \\ J_{2} (-A/\omega) \\ J_1 (-A/\omega) \\ J_0 (-A/\omega) \\ J_1 (-A/\omega) \\ J_2 (-A/\omega) \\ . \\ . \end{pmatrix}
 \end{align}
where $J_0 (A/\omega)$ is the $k^{\text{th}}$ and $k'^{\text{th}}$ component of the left and right eigenfunctions, whose energies are $\epsilon_0/2 + k\omega$ and $-\epsilon_0/2 + k' \omega$, respectively. Here $J_n$ is the $n^{\text{th}}$ Bessel function of the first kind. First, this implies that the width of wave-functions in harmonic space is given by $A/\omega$, as one may expect: the drive at any site can change the energy at most by $A$. Next, if we turn on hopping between these sites, and consider only resonant terms, the effective Hamiltonian governing the mixing of these left and right site eigenstates is given by

\begin{center}
\begin{figure}
\includegraphics[width=0.46 \textwidth]{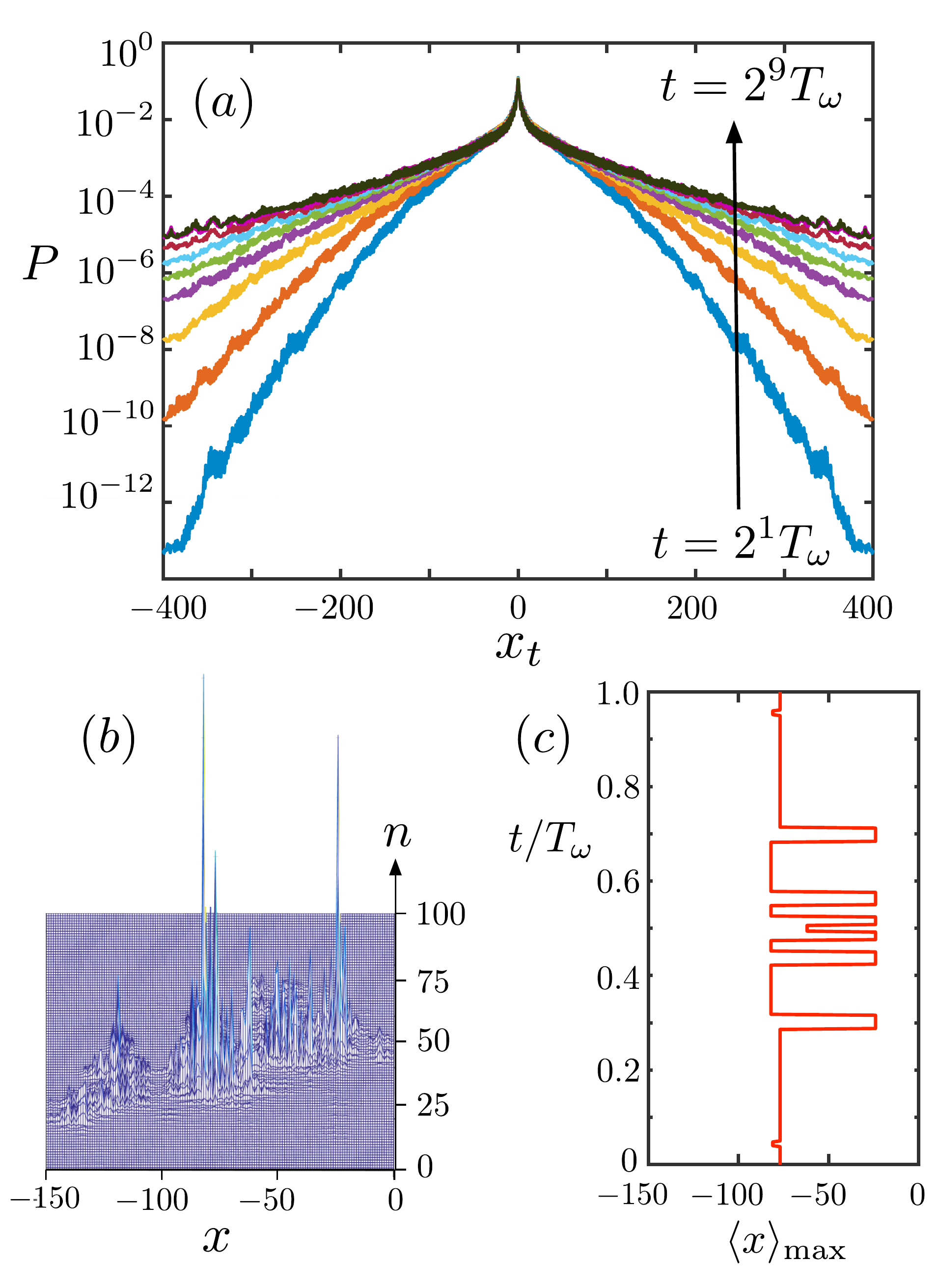}
\caption{(a) The full distribution $P(x_t | x_0 = 0)$ is plotted at times $t = 2^{[1, 2, 3, 4, 5, 6, 7, 8, 9]}$ $T_\omega$, for $W =1 $, $\omega = 0.02$, and $A = 0.2$. The probability distribution changes its slope logarithmically-in-time, which naturally gives rise to the logarithmic charge spread. Note that the disitrbution has weight at length scales much greater than $\xi_F \approx 75$ in this case, and illustrates the role of far-off resonances in the charge spread. (b) A characteristic Floquet eigenstate in real-space $x$ and harmonic space $n$; we see that real-space clusters are separated in harmonic space by $\sim 2A/\omega \approx 20$, as discussed in the main text. (c) The position of maximum probability $\avg{x}_{\text{max}}$ is plotted in this eigenstate over one time period $T_\omega$ and shows large LZ hops. }
\label{fig:dynextra}
\end{figure}
\end{center}

\begin{align}
H_{\text{eff.}} &= \begin{pmatrix} -\epsilon_0/2 &   \Delta J_{n_0} (2A/\omega)  \\ \Delta J_{n_0} (2A/\omega) & \epsilon_0/2 - n_0 \omega \end{pmatrix}
\label{eq:tls}
\end{align}

where we assume $\epsilon_0 \approx n_0 \omega$, and we used the relation $\sum_n J_{n} (x) J_{m-n} (x) = J_m (2 x)$. Assuming standard Anderson-model phenomenology applies, that is, $\Delta \sim W e^{-x/\xi}$, where $x$ represents the distance between the $L$ and $R$ sites, the condition for a resonance between these states becomes: $\omega \lesssim W e^{-x/\xi} J_{n_0} (A/\omega)$. Assuming we are in the strong driving regime, where $A \gg \omega$ (recall that $A_c$ depends only logarithmically on $\omega$, and thus, for $\omega \ll W$, $A_c > \omega$), the resonance condition is: $x < x_{\text{ad}} = \frac{\xi}{2} \text{ln} \left(\frac{W^2}{A \omega} \right)$, and $n_0 < A/\omega$. 

The second condition amounts to requiring that the energy difference between the two sites, $\epsilon_0$, is surmountable by the the change in local potential $\sim A$ due to the drive. Now, the probability for the second condition to be fulfilled between any two sites is $p = 1/x_{\text{LZ}} = A/W$. Concomitantly, within a radius $\sim x_{\text{LZ}}$ from the chosen site, we are certain to fulfill the second condition for a potential resonance. Finally, for the resonances to `proliferate', the necessary condition is $x_{\text{LZ}} < x_{\text{ad}}$. One can easily check that this occurs for a critical drive amplitude, $ A > A_c$ where $A_c$ scales as $\sim W$, $\text{ln} (\omega)$, as seen in the numerics.

The resonance structure discussed here is easily seen in eigenfunctions found numerically. A characteristic eigenfunction for $\omega = 0.02$, $A = 0.2$ has been plotted in Fig.~\ref{fig:dynextra} (b), where one clearly sees harmonic-space spread of $n \sim 2 A/\omega = 20$ of the eigenfunction at any on site. One also sees how clusters of real-space size $\sim 40$ emerge at the length $x_\text{ad}$ which in this case is a factor of 2 smaller than $\xi_F \sim 75$. The dynamics of the charge in a single time period $T_\omega$ in this eigenstate has been plotted in Fig.~\ref{fig:dynextra} (c) and shows clearly the LZ hopping.  

The above picture also brings out the quasi-1D nature of the problem. Thus, we again have an Anderson problem with renormalized hopping energies, which are disordered, and on-site energy mismatches of the order of $\omega$. Thus, we expect these resonances to interfere and localize. To make further predictions regarding the size of these Floquet eigenstates, and the dynamics, we consider a more intuitive picture of these resonances, in terms of adiabatic LZ transitions. 

\section{LZ orbits, exponentially slow dephasing, and weak localization.}

The role of LZ transitions in the heating, dynamics, and ultimate breakdown of the many-body-localized phase has already been considered in Refs.~\cite{abanin2016theory,heatingmbldrivenSG}. The resonances discussed above are also connected to LZ transitions in this single-particle situation. The probability of an adiabatic LZ transition between two sites during the course of the drive cycle is $P_{\text{ad}} = 1 - e^{- \Delta^2 / A \omega}$, where $\Delta = W e^{-x/\xi}$; we see that $P_{\text{ad}} \gtrsim 1/2$ for $x < x_{\text{ad}}$, as defined above. Moreover, a LZ crossing only occurs if the on-site energies are at most separated by energy $A$. This again implies that the probability of finding a level crossing with a chosen site within a distance $x_{\text{LZ}} = W/A$ is near unity. 

We can now use this picture to better describe the dynamics of the problem and understand the structure of Floquet eigenstates. We note that, as the local potential is varied, the first adiabatic LZ transition is encountered when the local potential changes by an amount $A' \approx W/x_{\text{ad}}$, and involves a transfer distance of $x_{\text{ad}}$. There are thus $\approx A/A' = x_{\text{ad}}/x_{\text{LZ}}$ such resonances in a given drive cycle. Thus, if the particle has a probability of hopping left or right, with roughly equal probability, it will end up traveling a distance $\xi_F \sim x_{\text{ad}} \left( x_{\text{ad}} / x_{\text{LZ}} \right)^{1/2}$ in a drive cycle. We denote $\xi_{F,\text{th}} = c (\omega, W) \xi_{F}$, where we set $c(\omega, W)$ such that $\xi_{F,\text{th}} (A_c) = \xi$ at $A = A_c (\omega, W)$, and the subscript in $\xi_{F,\text{th}}$ indicates that this is our theoretical expectation. Fig.~\ref{fig:xifscaling} shows that, in the strong driving regime, there is good agreement between $\xi_F$ found numerically and $\xi_{F,\text{th}}$.

When measured stroboscopically, at times $t = 2 \pi l /\omega$, charge motion occurs only due to hopping of the charge between these LZ orbits (which are themselves closed). Examining Eq.~(\ref{eq:tls}), we note that the time-scale at which this occurs is given by $\tau^{-1}_0 \sim W e^{-x/\xi_F} J_n (A/\omega)$, where $n$ is arbitrary and $x \gtrsim \xi_F$ is the distance between the centers of these orbits. The exponential decay is due to the locality of the Floquet eigenstates and the operator $\hat{x}$ at any given site. These couplings lead to effective Mott-like pairs~\cite{mottconductivity} between the LZ orbits which dephase at time scale $\tau_0$, resulting in charge $1/\xi_F$ to hop a distance $x$. The time it takes to transfer a unit of charge (a distance $x$), is thus $t (x) = \tau_0 \xi_F$; inverting this results in $x (t) \sim \xi_F \text{ln} \left( \text{const.} t \sqrt{\frac{\omega W^3}{A^2}} \right)$ where the $\text{const.}$ contains logarithmic dependencies on all parameters which we neglect. Fig.~\ref{fig:figdyn} (b) confirms the scaling predicted in this relation. 

We note that individual instances of charge motion from a given initial position do not show the ideal logarithmic behavior: this is because such dephasing occurs between widely separated LZ orbits, at distances $x \gg \xi_F$, and there are discrepancies between the spatial arrangment of different Floquet eigenstates in any given disorder realization. The role of long-distance hopping is clear from the fact that probability distribution $P(x_t |x_0)$ has non-zero weight at the longest scales $\sim L$ in the system, see Fig.~\ref{fig:dynextra} (a). We also note that, this logarithmic-in-time behavior must saturate eventually at a maximal distance $\sim \xi_F$; at infinite time, the particle has an exponentially small probability, in $x$, of being distance $x$ afar from the origin, and this leads to all moments of $P(x_t | x_0)$ being finite at infinite time.

The localization of Floquet eigenstates is clearly sensitive to the phase accumulated by the charge over the course of the complete drive cycle. One can check this by adding a random phase $\delta(x)$ at each site, as in Eq.~(\ref{eq:defhf}). As is clear from Fig.~\ref{fig:phasefig} This induces a flux $\Phi(x) = \delta(x) - \delta(x+1)$ in the plaquettes at site $x$ in the harmonic-space representation of the problem, and thus serves the role of a random magnetic field. As is characteristic of weak localization in two dimensions~\cite{leeramakrishnan}, we see that this tends to increase the localization length of the Floquet eigenstates, while preserving all other features of the dynamics. A more drastic change occurs for $\delta(x) = \text{const.}$. The Floquet Hamiltonian now corresponds to the Quantum Hall/Hofstadter~\cite{hofstadter} system and the energy gradient of $\omega$ in the harmonic space direction results in currents flowing in opposite directions along the ends of the harmonic-space `edges'. Due to disorder, charge can scatter between these edge states, and as a result, spreads diffusively.  

\section{Discussion and Conclusions}

In this work, we studied the strongly monochromatically driven Anderson model in one space dimension by examining the associated Floquet Hamiltonian in one higher harmonic-space dimension. We showed that resonances in this Hamiltonian correspond to adiabatic LZ transitions in the dynamical problem, and provided conditions for the proliferation of these resonances. In this strong driving regime, Floquet eigenstates can be understood as closed LZ orbits (of characteristic size and hops as discussed in the main text). Dephasing between faraway orbits forming Mott-like pairs occurs at exponentially slow time-scales and gives rise to logarithmic-in-time charge spread. The orbits themselves are constructed by interference effects which are consequently affected by randomizing the local phase of the drive. A more systematic study of the dynamics in the presence of random drive phases is left for future work.

We note that our arguments likely apply as is to Anderson insulators in two dimensions. However, the presence of a mobility edge in three dimensions and greater~\cite{scalingtheoryandersonlocalization} is likely to destroy the localization effects we study here. A more numerically accessible test of the effect of a mobility edge would lie in studying the one-dimensional Aubrey-Andre model~\cite{aubry1980analyticity} and its generalizations~\cite{ganeshansingleparticlemobilityedge}. We note in passing that multiple incommensurate-frequency drives can be studied as higher dimensional (one dimension for each drive) analogues of the problem we have considered here. By analogy with known results from the Anderson model in $d$-dimensions, we expect the system to become delocalized upon the application of three drive frequencies; this is surprising since two incommensurate frequencies are enough to form a `bath' that is dense in frequency space. It is also worth asking whether logarithmic-in-time behavior seen in the energy spread, as seen~\cite{rehnheatingmbl} at the intersection of the Floquet-ergodic and Floquet-MBL phases occurs via a phase locking of LZ transitions between many-body eigenstates in a manner similar to the real-space LZ orbits found here. 

\section{Acknowledgements}

We thank Sarang Gopalakrishnan and David Huse for useful discussions. KA and RNB acknowledge support from DOE-BES Grant DE-SC0002140 (RNB) and U.K. Foundation (KA).


\begin{thebibliography}{63}%
\makeatletter
\providecommand \@ifxundefined [1]{%
 \@ifx{#1\undefined}
}%
\providecommand \@ifnum [1]{%
 \ifnum #1\expandafter \@firstoftwo
 \else \expandafter \@secondoftwo
 \fi
}%
\providecommand \@ifx [1]{%
 \ifx #1\expandafter \@firstoftwo
 \else \expandafter \@secondoftwo
 \fi
}%
\providecommand \natexlab [1]{#1}%
\providecommand \enquote  [1]{``#1''}%
\providecommand \bibnamefont  [1]{#1}%
\providecommand \bibfnamefont [1]{#1}%
\providecommand \citenamefont [1]{#1}%
\providecommand \href@noop [0]{\@secondoftwo}%
\providecommand \href [0]{\begingroup \@sanitize@url \@href}%
\providecommand \@href[1]{\@@startlink{#1}\@@href}%
\providecommand \@@href[1]{\endgroup#1\@@endlink}%
\providecommand \@sanitize@url [0]{\catcode `\\12\catcode `\$12\catcode
  `\&12\catcode `\#12\catcode `\^12\catcode `\_12\catcode `\%12\relax}%
\providecommand \@@startlink[1]{}%
\providecommand \@@endlink[0]{}%
\providecommand \url  [0]{\begingroup\@sanitize@url \@url }%
\providecommand \@url [1]{\endgroup\@href {#1}{\urlprefix }}%
\providecommand \urlprefix  [0]{URL }%
\providecommand \Eprint [0]{\href }%
\providecommand \doibase [0]{http://dx.doi.org/}%
\providecommand \selectlanguage [0]{\@gobble}%
\providecommand \bibinfo  [0]{\@secondoftwo}%
\providecommand \bibfield  [0]{\@secondoftwo}%
\providecommand \translation [1]{[#1]}%
\providecommand \BibitemOpen [0]{}%
\providecommand \bibitemStop [0]{}%
\providecommand \bibitemNoStop [0]{.\EOS\space}%
\providecommand \EOS [0]{\spacefactor3000\relax}%
\providecommand \BibitemShut  [1]{\csname bibitem#1\endcsname}%
\let\auto@bib@innerbib\@empty
\bibitem [{\citenamefont {Greiner}\ \emph {et~al.}(2002)\citenamefont
  {Greiner}, \citenamefont {Mandel}, \citenamefont {Hasch},\ and\ \citenamefont
  {Bloch}}]{Greiner}%
  \BibitemOpen
  \bibfield  {author} {\bibinfo {author} {\bibfnamefont {M.}~\bibnamefont
  {Greiner}}, \bibinfo {author} {\bibfnamefont {O.}~\bibnamefont {Mandel}},
  \bibinfo {author} {\bibfnamefont {T.~W.}\ \bibnamefont {Hasch}}, \ and\
  \bibinfo {author} {\bibfnamefont {I.}~\bibnamefont {Bloch}},\ }\href@noop {}
  {\bibfield  {journal} {\bibinfo  {journal} {Nature}\ }\textbf {\bibinfo
  {volume} {419}},\ \bibinfo {pages} {51} (\bibinfo {year} {2002})}\BibitemShut
  {NoStop}%
\bibitem [{\citenamefont {Kinoshita}\ \emph {et~al.}(2006)\citenamefont
  {Kinoshita}, \citenamefont {Wenger},\ and\ \citenamefont
  {Weiss}}]{Kinoshita06}%
  \BibitemOpen
  \bibfield  {author} {\bibinfo {author} {\bibfnamefont {T.}~\bibnamefont
  {Kinoshita}}, \bibinfo {author} {\bibfnamefont {T.}~\bibnamefont {Wenger}}, \
  and\ \bibinfo {author} {\bibfnamefont {D.~S.}\ \bibnamefont {Weiss}},\
  }\href@noop {} {\bibfield  {journal} {\bibinfo  {journal} {Nature}\ }\textbf
  {\bibinfo {volume} {440}},\ \bibinfo {pages} {900} (\bibinfo {year}
  {2006})}\BibitemShut {NoStop}%
\bibitem [{\citenamefont {Langen}\ \emph {et~al.}(2013)\citenamefont {Langen},
  \citenamefont {Geiger}, \citenamefont {Kuhnert}, \citenamefont {Rauer},\ and\
  \citenamefont {Schmiedmayer}}]{Langen}%
  \BibitemOpen
  \bibfield  {author} {\bibinfo {author} {\bibfnamefont {T.}~\bibnamefont
  {Langen}}, \bibinfo {author} {\bibfnamefont {R.}~\bibnamefont {Geiger}},
  \bibinfo {author} {\bibfnamefont {M.}~\bibnamefont {Kuhnert}}, \bibinfo
  {author} {\bibfnamefont {B.}~\bibnamefont {Rauer}}, \ and\ \bibinfo {author}
  {\bibfnamefont {J.}~\bibnamefont {Schmiedmayer}},\ }\href@noop {} {\bibfield
  {journal} {\bibinfo  {journal} {Nature Physics}\ }\textbf {\bibinfo {volume}
  {9}},\ \bibinfo {pages} {640} (\bibinfo {year} {2013})}\BibitemShut {NoStop}%
\bibitem [{\citenamefont {Schreiber}\ \emph {et~al.}(2015)\citenamefont
  {Schreiber}, \citenamefont {Hodgman}, \citenamefont {Bordia}, \citenamefont
  {L{\"u}schen}, \citenamefont {Fischer}, \citenamefont {Vosk}, \citenamefont
  {Altman}, \citenamefont {Schneider},\ and\ \citenamefont
  {Bloch}}]{schreiber2015observation}%
  \BibitemOpen
  \bibfield  {author} {\bibinfo {author} {\bibfnamefont {M.}~\bibnamefont
  {Schreiber}}, \bibinfo {author} {\bibfnamefont {S.~S.}\ \bibnamefont
  {Hodgman}}, \bibinfo {author} {\bibfnamefont {P.}~\bibnamefont {Bordia}},
  \bibinfo {author} {\bibfnamefont {H.~P.}\ \bibnamefont {L{\"u}schen}},
  \bibinfo {author} {\bibfnamefont {M.~H.}\ \bibnamefont {Fischer}}, \bibinfo
  {author} {\bibfnamefont {R.}~\bibnamefont {Vosk}}, \bibinfo {author}
  {\bibfnamefont {E.}~\bibnamefont {Altman}}, \bibinfo {author} {\bibfnamefont
  {U.}~\bibnamefont {Schneider}}, \ and\ \bibinfo {author} {\bibfnamefont
  {I.}~\bibnamefont {Bloch}},\ }\href@noop {} {\bibfield  {journal} {\bibinfo
  {journal} {Science}\ }\textbf {\bibinfo {volume} {349}},\ \bibinfo {pages}
  {842} (\bibinfo {year} {2015})}\BibitemShut {NoStop}%
\bibitem [{\citenamefont {Smith}\ \emph {et~al.}(2016)\citenamefont {Smith},
  \citenamefont {Lee}, \citenamefont {Richerme}, \citenamefont {Neyenhuis},
  \citenamefont {Hess}, \citenamefont {Hauke}, \citenamefont {Heyl},
  \citenamefont {Huse},\ and\ \citenamefont {Monroe}}]{smith2016many}%
  \BibitemOpen
  \bibfield  {author} {\bibinfo {author} {\bibfnamefont {J.}~\bibnamefont
  {Smith}}, \bibinfo {author} {\bibfnamefont {A.}~\bibnamefont {Lee}}, \bibinfo
  {author} {\bibfnamefont {P.}~\bibnamefont {Richerme}}, \bibinfo {author}
  {\bibfnamefont {B.}~\bibnamefont {Neyenhuis}}, \bibinfo {author}
  {\bibfnamefont {P.~W.}\ \bibnamefont {Hess}}, \bibinfo {author}
  {\bibfnamefont {P.}~\bibnamefont {Hauke}}, \bibinfo {author} {\bibfnamefont
  {M.}~\bibnamefont {Heyl}}, \bibinfo {author} {\bibfnamefont {D.~A.}\
  \bibnamefont {Huse}}, \ and\ \bibinfo {author} {\bibfnamefont
  {C.}~\bibnamefont {Monroe}},\ }\href@noop {} {\bibfield  {journal} {\bibinfo
  {journal} {Nature Physics}\ }\textbf {\bibinfo {volume} {12}},\ \bibinfo
  {pages} {907} (\bibinfo {year} {2016})}\BibitemShut {NoStop}%
\bibitem [{\citenamefont {Cheneau}\ \emph {et~al.}(2012)\citenamefont
  {Cheneau}, \citenamefont {Barmettler}, \citenamefont {Poletti}, \citenamefont
  {Endres}, \citenamefont {Schau{\ss}}, \citenamefont {Fukuhara}, \citenamefont
  {Gross}, \citenamefont {Bloch}, \citenamefont {Kollath},\ and\ \citenamefont
  {Kuhr}}]{Cheneau}%
  \BibitemOpen
  \bibfield  {author} {\bibinfo {author} {\bibfnamefont {M.}~\bibnamefont
  {Cheneau}}, \bibinfo {author} {\bibfnamefont {P.}~\bibnamefont {Barmettler}},
  \bibinfo {author} {\bibfnamefont {D.}~\bibnamefont {Poletti}}, \bibinfo
  {author} {\bibfnamefont {M.}~\bibnamefont {Endres}}, \bibinfo {author}
  {\bibfnamefont {P.}~\bibnamefont {Schau{\ss}}}, \bibinfo {author}
  {\bibfnamefont {T.}~\bibnamefont {Fukuhara}}, \bibinfo {author}
  {\bibfnamefont {C.}~\bibnamefont {Gross}}, \bibinfo {author} {\bibfnamefont
  {I.}~\bibnamefont {Bloch}}, \bibinfo {author} {\bibfnamefont
  {C.}~\bibnamefont {Kollath}}, \ and\ \bibinfo {author} {\bibfnamefont
  {S.}~\bibnamefont {Kuhr}},\ }\href@noop {} {\bibfield  {journal} {\bibinfo
  {journal} {Nature}\ }\textbf {\bibinfo {volume} {481}},\ \bibinfo {pages}
  {484} (\bibinfo {year} {2012})}\BibitemShut {NoStop}%
\bibitem [{\citenamefont {Sadler}\ \emph {et~al.}(2006)\citenamefont {Sadler},
  \citenamefont {Higbie}, \citenamefont {Leslie}, \citenamefont
  {Vengalattore},\ and\ \citenamefont {Stamper-Kurn}}]{sadler2006spontaneous}%
  \BibitemOpen
  \bibfield  {author} {\bibinfo {author} {\bibfnamefont {L.}~\bibnamefont
  {Sadler}}, \bibinfo {author} {\bibfnamefont {J.}~\bibnamefont {Higbie}},
  \bibinfo {author} {\bibfnamefont {S.}~\bibnamefont {Leslie}}, \bibinfo
  {author} {\bibfnamefont {M.}~\bibnamefont {Vengalattore}}, \ and\ \bibinfo
  {author} {\bibfnamefont {D.}~\bibnamefont {Stamper-Kurn}},\ }\href@noop {}
  {\bibfield  {journal} {\bibinfo  {journal} {Nature}\ }\textbf {\bibinfo
  {volume} {443}},\ \bibinfo {pages} {312} (\bibinfo {year}
  {2006})}\BibitemShut {NoStop}%
\bibitem [{\citenamefont {Cooper}(2011)}]{cooperfluxlattice}%
  \BibitemOpen
  \bibfield  {author} {\bibinfo {author} {\bibfnamefont {N.~R.}\ \bibnamefont
  {Cooper}},\ }\href {\doibase 10.1103/PhysRevLett.106.175301} {\bibfield
  {journal} {\bibinfo  {journal} {Phys. Rev. Lett.}\ }\textbf {\bibinfo
  {volume} {106}},\ \bibinfo {pages} {175301} (\bibinfo {year}
  {2011})}\BibitemShut {NoStop}%
\bibitem [{\citenamefont {Kitagawa}\ \emph {et~al.}(2011)\citenamefont
  {Kitagawa}, \citenamefont {Oka}, \citenamefont {Brataas}, \citenamefont
  {Fu},\ and\ \citenamefont {Demler}}]{kitagawa2011transport}%
  \BibitemOpen
  \bibfield  {author} {\bibinfo {author} {\bibfnamefont {T.}~\bibnamefont
  {Kitagawa}}, \bibinfo {author} {\bibfnamefont {T.}~\bibnamefont {Oka}},
  \bibinfo {author} {\bibfnamefont {A.}~\bibnamefont {Brataas}}, \bibinfo
  {author} {\bibfnamefont {L.}~\bibnamefont {Fu}}, \ and\ \bibinfo {author}
  {\bibfnamefont {E.}~\bibnamefont {Demler}},\ }\href@noop {} {\bibfield
  {journal} {\bibinfo  {journal} {Physical Review B}\ }\textbf {\bibinfo
  {volume} {84}},\ \bibinfo {pages} {235108} (\bibinfo {year}
  {2011})}\BibitemShut {NoStop}%
\bibitem [{\citenamefont {Galitski}\ and\ \citenamefont
  {Spielman}(2013)}]{galitski2013spin}%
  \BibitemOpen
  \bibfield  {author} {\bibinfo {author} {\bibfnamefont {V.}~\bibnamefont
  {Galitski}}\ and\ \bibinfo {author} {\bibfnamefont {I.~B.}\ \bibnamefont
  {Spielman}},\ }\href@noop {} {\bibfield  {journal} {\bibinfo  {journal}
  {Nature}\ }\textbf {\bibinfo {volume} {494}},\ \bibinfo {pages} {49}
  (\bibinfo {year} {2013})}\BibitemShut {NoStop}%
\bibitem [{\citenamefont {Dalibard}\ \emph {et~al.}(2011)\citenamefont
  {Dalibard}, \citenamefont {Gerbier}, \citenamefont
  {Juzeli\ifmmode~\bar{u}\else \={u}\fi{}nas},\ and\ \citenamefont
  {\"Ohberg}}]{dalibardguagepotentials}%
  \BibitemOpen
  \bibfield  {author} {\bibinfo {author} {\bibfnamefont {J.}~\bibnamefont
  {Dalibard}}, \bibinfo {author} {\bibfnamefont {F.}~\bibnamefont {Gerbier}},
  \bibinfo {author} {\bibfnamefont {G.}~\bibnamefont
  {Juzeli\ifmmode~\bar{u}\else \={u}\fi{}nas}}, \ and\ \bibinfo {author}
  {\bibfnamefont {P.}~\bibnamefont {\"Ohberg}},\ }\href {\doibase
  10.1103/RevModPhys.83.1523} {\bibfield  {journal} {\bibinfo  {journal} {Rev.
  Mod. Phys.}\ }\textbf {\bibinfo {volume} {83}},\ \bibinfo {pages} {1523}
  (\bibinfo {year} {2011})}\BibitemShut {NoStop}%
\bibitem [{\citenamefont {Kennedy}\ \emph {et~al.}(2015)\citenamefont
  {Kennedy}, \citenamefont {Burton}, \citenamefont {Chung},\ and\ \citenamefont
  {Ketterle}}]{kennedy2015observation}%
  \BibitemOpen
  \bibfield  {author} {\bibinfo {author} {\bibfnamefont {C.~J.}\ \bibnamefont
  {Kennedy}}, \bibinfo {author} {\bibfnamefont {W.~C.}\ \bibnamefont {Burton}},
  \bibinfo {author} {\bibfnamefont {W.~C.}\ \bibnamefont {Chung}}, \ and\
  \bibinfo {author} {\bibfnamefont {W.}~\bibnamefont {Ketterle}},\ }\href@noop
  {} {\bibfield  {journal} {\bibinfo  {journal} {Nature Physics}\ }\textbf
  {\bibinfo {volume} {11}},\ \bibinfo {pages} {859} (\bibinfo {year}
  {2015})}\BibitemShut {NoStop}%
\bibitem [{\citenamefont {D'Alessio}\ and\ \citenamefont
  {Rigol}(2014)}]{LucaIsolatedDriven}%
  \BibitemOpen
  \bibfield  {author} {\bibinfo {author} {\bibfnamefont {L.}~\bibnamefont
  {D'Alessio}}\ and\ \bibinfo {author} {\bibfnamefont {M.}~\bibnamefont
  {Rigol}},\ }\href {\doibase 10.1103/PhysRevX.4.041048} {\bibfield  {journal}
  {\bibinfo  {journal} {Phys. Rev. X}\ }\textbf {\bibinfo {volume} {4}},\
  \bibinfo {pages} {041048} (\bibinfo {year} {2014})}\BibitemShut {NoStop}%
\bibitem [{\citenamefont {D?Alessio}\ and\ \citenamefont
  {Polkovnikov}(2013)}]{dalessio2013heating}%
  \BibitemOpen
  \bibfield  {author} {\bibinfo {author} {\bibfnamefont {L.}~\bibnamefont
  {D?Alessio}}\ and\ \bibinfo {author} {\bibfnamefont {A.}~\bibnamefont
  {Polkovnikov}},\ }\href@noop {} {\bibfield  {journal} {\bibinfo  {journal}
  {Annals of Physics}\ }\textbf {\bibinfo {volume} {333}},\ \bibinfo {pages}
  {19} (\bibinfo {year} {2013})}\BibitemShut {NoStop}%
\bibitem [{\citenamefont {Ponte}\ \emph
  {et~al.}(2015{\natexlab{a}})\citenamefont {Ponte}, \citenamefont {Chandran},
  \citenamefont {Papi{\'c}},\ and\ \citenamefont
  {Abanin}}]{ponte2015periodically}%
  \BibitemOpen
  \bibfield  {author} {\bibinfo {author} {\bibfnamefont {P.}~\bibnamefont
  {Ponte}}, \bibinfo {author} {\bibfnamefont {A.}~\bibnamefont {Chandran}},
  \bibinfo {author} {\bibfnamefont {Z.}~\bibnamefont {Papi{\'c}}}, \ and\
  \bibinfo {author} {\bibfnamefont {D.~A.}\ \bibnamefont {Abanin}},\
  }\href@noop {} {\bibfield  {journal} {\bibinfo  {journal} {Annals of
  Physics}\ }\textbf {\bibinfo {volume} {353}},\ \bibinfo {pages} {196}
  (\bibinfo {year} {2015}{\natexlab{a}})}\BibitemShut {NoStop}%
\bibitem [{\citenamefont {Rehn}\ \emph {et~al.}(2016)\citenamefont {Rehn},
  \citenamefont {Lazarides}, \citenamefont {Pollmann},\ and\ \citenamefont
  {Moessner}}]{rehnheatingmbl}%
  \BibitemOpen
  \bibfield  {author} {\bibinfo {author} {\bibfnamefont {J.}~\bibnamefont
  {Rehn}}, \bibinfo {author} {\bibfnamefont {A.}~\bibnamefont {Lazarides}},
  \bibinfo {author} {\bibfnamefont {F.}~\bibnamefont {Pollmann}}, \ and\
  \bibinfo {author} {\bibfnamefont {R.}~\bibnamefont {Moessner}},\ }\href
  {\doibase 10.1103/PhysRevB.94.020201} {\bibfield  {journal} {\bibinfo
  {journal} {Phys. Rev. B}\ }\textbf {\bibinfo {volume} {94}},\ \bibinfo
  {pages} {020201} (\bibinfo {year} {2016})}\BibitemShut {NoStop}%
\bibitem [{\citenamefont {Bukov}\ \emph {et~al.}(2015)\citenamefont {Bukov},
  \citenamefont {Gopalakrishnan}, \citenamefont {Knap},\ and\ \citenamefont
  {Demler}}]{prethermallonglived}%
  \BibitemOpen
  \bibfield  {author} {\bibinfo {author} {\bibfnamefont {M.}~\bibnamefont
  {Bukov}}, \bibinfo {author} {\bibfnamefont {S.}~\bibnamefont
  {Gopalakrishnan}}, \bibinfo {author} {\bibfnamefont {M.}~\bibnamefont
  {Knap}}, \ and\ \bibinfo {author} {\bibfnamefont {E.}~\bibnamefont
  {Demler}},\ }\href {\doibase 10.1103/PhysRevLett.115.205301} {\bibfield
  {journal} {\bibinfo  {journal} {Phys. Rev. Lett.}\ }\textbf {\bibinfo
  {volume} {115}},\ \bibinfo {pages} {205301} (\bibinfo {year}
  {2015})}\BibitemShut {NoStop}%
\bibitem [{\citenamefont {Abanin}\ \emph
  {et~al.}(2015{\natexlab{a}})\citenamefont {Abanin}, \citenamefont {De~Roeck},
  \citenamefont {Ho},\ and\ \citenamefont {Huveneers}}]{abanin2015effective}%
  \BibitemOpen
  \bibfield  {author} {\bibinfo {author} {\bibfnamefont {D.~A.}\ \bibnamefont
  {Abanin}}, \bibinfo {author} {\bibfnamefont {W.}~\bibnamefont {De~Roeck}},
  \bibinfo {author} {\bibfnamefont {W.~W.}\ \bibnamefont {Ho}}, \ and\ \bibinfo
  {author} {\bibfnamefont {F.}~\bibnamefont {Huveneers}},\ }\href@noop {}
  {\bibfield  {journal} {\bibinfo  {journal} {arXiv preprint arXiv:1510.03405}\
  } (\bibinfo {year} {2015}{\natexlab{a}})}\BibitemShut {NoStop}%
\bibitem [{\citenamefont {Abanin}\ \emph
  {et~al.}(2015{\natexlab{b}})\citenamefont {Abanin}, \citenamefont {De~Roeck},
  \citenamefont {Huveneers},\ and\ \citenamefont {Ho}}]{abanin2015rigorous}%
  \BibitemOpen
  \bibfield  {author} {\bibinfo {author} {\bibfnamefont {D.}~\bibnamefont
  {Abanin}}, \bibinfo {author} {\bibfnamefont {W.}~\bibnamefont {De~Roeck}},
  \bibinfo {author} {\bibfnamefont {F.}~\bibnamefont {Huveneers}}, \ and\
  \bibinfo {author} {\bibfnamefont {W.~W.}\ \bibnamefont {Ho}},\ }\href@noop {}
  {\bibfield  {journal} {\bibinfo  {journal} {arXiv preprint arXiv:1509.05386}\
  } (\bibinfo {year} {2015}{\natexlab{b}})}\BibitemShut {NoStop}%
\bibitem [{\citenamefont {Bukov}\ \emph {et~al.}(2016)\citenamefont {Bukov},
  \citenamefont {Heyl}, \citenamefont {Huse},\ and\ \citenamefont
  {Polkovnikov}}]{bukovtwoband}%
  \BibitemOpen
  \bibfield  {author} {\bibinfo {author} {\bibfnamefont {M.}~\bibnamefont
  {Bukov}}, \bibinfo {author} {\bibfnamefont {M.}~\bibnamefont {Heyl}},
  \bibinfo {author} {\bibfnamefont {D.~A.}\ \bibnamefont {Huse}}, \ and\
  \bibinfo {author} {\bibfnamefont {A.}~\bibnamefont {Polkovnikov}},\ }\href
  {\doibase 10.1103/PhysRevB.93.155132} {\bibfield  {journal} {\bibinfo
  {journal} {Phys. Rev. B}\ }\textbf {\bibinfo {volume} {93}},\ \bibinfo
  {pages} {155132} (\bibinfo {year} {2016})}\BibitemShut {NoStop}%
\bibitem [{\citenamefont {Ponte}\ \emph
  {et~al.}(2015{\natexlab{b}})\citenamefont {Ponte}, \citenamefont
  {Papi\ifmmode~\acute{c}\else \'{c}\fi{}}, \citenamefont {Huveneers},\ and\
  \citenamefont {Abanin}}]{pontefloquetMBL}%
  \BibitemOpen
  \bibfield  {author} {\bibinfo {author} {\bibfnamefont {P.}~\bibnamefont
  {Ponte}}, \bibinfo {author} {\bibfnamefont {Z.}~\bibnamefont
  {Papi\ifmmode~\acute{c}\else \'{c}\fi{}}}, \bibinfo {author} {\bibfnamefont
  {F.~m.~c.}\ \bibnamefont {Huveneers}}, \ and\ \bibinfo {author}
  {\bibfnamefont {D.~A.}\ \bibnamefont {Abanin}},\ }\href {\doibase
  10.1103/PhysRevLett.114.140401} {\bibfield  {journal} {\bibinfo  {journal}
  {Phys. Rev. Lett.}\ }\textbf {\bibinfo {volume} {114}},\ \bibinfo {pages}
  {140401} (\bibinfo {year} {2015}{\natexlab{b}})}\BibitemShut {NoStop}%
\bibitem [{\citenamefont {Lazarides}\ \emph {et~al.}(2015)\citenamefont
  {Lazarides}, \citenamefont {Das},\ and\ \citenamefont
  {Moessner}}]{lazaridesperiodicmbl}%
  \BibitemOpen
  \bibfield  {author} {\bibinfo {author} {\bibfnamefont {A.}~\bibnamefont
  {Lazarides}}, \bibinfo {author} {\bibfnamefont {A.}~\bibnamefont {Das}}, \
  and\ \bibinfo {author} {\bibfnamefont {R.}~\bibnamefont {Moessner}},\ }\href
  {\doibase 10.1103/PhysRevLett.115.030402} {\bibfield  {journal} {\bibinfo
  {journal} {Phys. Rev. Lett.}\ }\textbf {\bibinfo {volume} {115}},\ \bibinfo
  {pages} {030402} (\bibinfo {year} {2015})}\BibitemShut {NoStop}%
\bibitem [{\citenamefont {Abanin}\ \emph {et~al.}(2016)\citenamefont {Abanin},
  \citenamefont {De~Roeck},\ and\ \citenamefont
  {Huveneers}}]{abanin2016theory}%
  \BibitemOpen
  \bibfield  {author} {\bibinfo {author} {\bibfnamefont {D.~A.}\ \bibnamefont
  {Abanin}}, \bibinfo {author} {\bibfnamefont {W.}~\bibnamefont {De~Roeck}}, \
  and\ \bibinfo {author} {\bibfnamefont {F.}~\bibnamefont {Huveneers}},\
  }\href@noop {} {\bibfield  {journal} {\bibinfo  {journal} {Annals of
  Physics}\ }\textbf {\bibinfo {volume} {372}},\ \bibinfo {pages} {1} (\bibinfo
  {year} {2016})}\BibitemShut {NoStop}%
\bibitem [{\citenamefont {Khemani}\ \emph {et~al.}(2016)\citenamefont
  {Khemani}, \citenamefont {Lazarides}, \citenamefont {Moessner},\ and\
  \citenamefont {Sondhi}}]{vedikatimecrystal}%
  \BibitemOpen
  \bibfield  {author} {\bibinfo {author} {\bibfnamefont {V.}~\bibnamefont
  {Khemani}}, \bibinfo {author} {\bibfnamefont {A.}~\bibnamefont {Lazarides}},
  \bibinfo {author} {\bibfnamefont {R.}~\bibnamefont {Moessner}}, \ and\
  \bibinfo {author} {\bibfnamefont {S.~L.}\ \bibnamefont {Sondhi}},\ }\href
  {\doibase 10.1103/PhysRevLett.116.250401} {\bibfield  {journal} {\bibinfo
  {journal} {Phys. Rev. Lett.}\ }\textbf {\bibinfo {volume} {116}},\ \bibinfo
  {pages} {250401} (\bibinfo {year} {2016})}\BibitemShut {NoStop}%
\bibitem [{\citenamefont {Else}\ \emph {et~al.}(2016)\citenamefont {Else},
  \citenamefont {Bauer},\ and\ \citenamefont {Nayak}}]{elsetimecrystal}%
  \BibitemOpen
  \bibfield  {author} {\bibinfo {author} {\bibfnamefont {D.~V.}\ \bibnamefont
  {Else}}, \bibinfo {author} {\bibfnamefont {B.}~\bibnamefont {Bauer}}, \ and\
  \bibinfo {author} {\bibfnamefont {C.}~\bibnamefont {Nayak}},\ }\href
  {\doibase 10.1103/PhysRevLett.117.090402} {\bibfield  {journal} {\bibinfo
  {journal} {Phys. Rev. Lett.}\ }\textbf {\bibinfo {volume} {117}},\ \bibinfo
  {pages} {090402} (\bibinfo {year} {2016})}\BibitemShut {NoStop}%
\bibitem [{\citenamefont {von Keyserlingk}\ \emph {et~al.}(2016)\citenamefont
  {von Keyserlingk}, \citenamefont {Khemani},\ and\ \citenamefont
  {Sondhi}}]{keyserlingktimecrystalstability}%
  \BibitemOpen
  \bibfield  {author} {\bibinfo {author} {\bibfnamefont {C.~W.}\ \bibnamefont
  {von Keyserlingk}}, \bibinfo {author} {\bibfnamefont {V.}~\bibnamefont
  {Khemani}}, \ and\ \bibinfo {author} {\bibfnamefont {S.~L.}\ \bibnamefont
  {Sondhi}},\ }\href {\doibase 10.1103/PhysRevB.94.085112} {\bibfield
  {journal} {\bibinfo  {journal} {Phys. Rev. B}\ }\textbf {\bibinfo {volume}
  {94}},\ \bibinfo {pages} {085112} (\bibinfo {year} {2016})}\BibitemShut
  {NoStop}%
\bibitem [{\citenamefont {Yao}\ \emph {et~al.}(2017)\citenamefont {Yao},
  \citenamefont {Potter}, \citenamefont {Potirniche},\ and\ \citenamefont
  {Vishwanath}}]{YaoRigidity}%
  \BibitemOpen
  \bibfield  {author} {\bibinfo {author} {\bibfnamefont {N.~Y.}\ \bibnamefont
  {Yao}}, \bibinfo {author} {\bibfnamefont {A.~C.}\ \bibnamefont {Potter}},
  \bibinfo {author} {\bibfnamefont {I.-D.}\ \bibnamefont {Potirniche}}, \ and\
  \bibinfo {author} {\bibfnamefont {A.}~\bibnamefont {Vishwanath}},\ }\href
  {\doibase 10.1103/PhysRevLett.118.030401} {\bibfield  {journal} {\bibinfo
  {journal} {Phys. Rev. Lett.}\ }\textbf {\bibinfo {volume} {118}},\ \bibinfo
  {pages} {030401} (\bibinfo {year} {2017})}\BibitemShut {NoStop}%
\bibitem [{\citenamefont {Bairey}\ \emph {et~al.}(2017)\citenamefont {Bairey},
  \citenamefont {Refael},\ and\ \citenamefont {Lindner}}]{bairey2017driving}%
  \BibitemOpen
  \bibfield  {author} {\bibinfo {author} {\bibfnamefont {E.}~\bibnamefont
  {Bairey}}, \bibinfo {author} {\bibfnamefont {G.}~\bibnamefont {Refael}}, \
  and\ \bibinfo {author} {\bibfnamefont {N.~H.}\ \bibnamefont {Lindner}},\
  }\href@noop {} {\bibfield  {journal} {\bibinfo  {journal} {arXiv preprint
  arXiv:1702.06208}\ } (\bibinfo {year} {2017})}\BibitemShut {NoStop}%
\bibitem [{\citenamefont {Ho}\ \emph {et~al.}(2017)\citenamefont {Ho},
  \citenamefont {Choi}, \citenamefont {Lukin},\ and\ \citenamefont
  {Abanin}}]{ho2017critical}%
  \BibitemOpen
  \bibfield  {author} {\bibinfo {author} {\bibfnamefont {W.~W.}\ \bibnamefont
  {Ho}}, \bibinfo {author} {\bibfnamefont {S.}~\bibnamefont {Choi}}, \bibinfo
  {author} {\bibfnamefont {M.~D.}\ \bibnamefont {Lukin}}, \ and\ \bibinfo
  {author} {\bibfnamefont {D.~A.}\ \bibnamefont {Abanin}},\ }\href@noop {}
  {\bibfield  {journal} {\bibinfo  {journal} {arXiv preprint arXiv:1703.04593}\
  } (\bibinfo {year} {2017})}\BibitemShut {NoStop}%
\bibitem [{\citenamefont {Gritsev}\ and\ \citenamefont
  {Polkovnikov}(2017)}]{gritsev2017integrable}%
  \BibitemOpen
  \bibfield  {author} {\bibinfo {author} {\bibfnamefont {V.}~\bibnamefont
  {Gritsev}}\ and\ \bibinfo {author} {\bibfnamefont {A.}~\bibnamefont
  {Polkovnikov}},\ }\href@noop {} {\bibfield  {journal} {\bibinfo  {journal}
  {arXiv preprint arXiv:1701.05276}\ } (\bibinfo {year} {2017})}\BibitemShut
  {NoStop}%
\bibitem [{\citenamefont {Martin}\ \emph {et~al.}(2016)\citenamefont {Martin},
  \citenamefont {Refael},\ and\ \citenamefont
  {Halperin}}]{martin2016topological}%
  \BibitemOpen
  \bibfield  {author} {\bibinfo {author} {\bibfnamefont {I.}~\bibnamefont
  {Martin}}, \bibinfo {author} {\bibfnamefont {G.}~\bibnamefont {Refael}}, \
  and\ \bibinfo {author} {\bibfnamefont {B.}~\bibnamefont {Halperin}},\
  }\href@noop {} {\bibfield  {journal} {\bibinfo  {journal} {arXiv preprint
  arXiv:1612.02143}\ } (\bibinfo {year} {2016})}\BibitemShut {NoStop}%
\bibitem [{\citenamefont {Potter}\ \emph {et~al.}(2016)\citenamefont {Potter},
  \citenamefont {Morimoto},\ and\ \citenamefont
  {Vishwanath}}]{potter2016topological}%
  \BibitemOpen
  \bibfield  {author} {\bibinfo {author} {\bibfnamefont {A.~C.}\ \bibnamefont
  {Potter}}, \bibinfo {author} {\bibfnamefont {T.}~\bibnamefont {Morimoto}}, \
  and\ \bibinfo {author} {\bibfnamefont {A.}~\bibnamefont {Vishwanath}},\
  }\href@noop {} {\bibfield  {journal} {\bibinfo  {journal} {arXiv preprint
  arXiv:1602.05194}\ } (\bibinfo {year} {2016})}\BibitemShut {NoStop}%
\bibitem [{\citenamefont {Po}\ \emph {et~al.}(2016)\citenamefont {Po},
  \citenamefont {Fidkowski}, \citenamefont {Morimoto}, \citenamefont {Potter},\
  and\ \citenamefont {Vishwanath}}]{Pochiralfloquetbosons}%
  \BibitemOpen
  \bibfield  {author} {\bibinfo {author} {\bibfnamefont {H.~C.}\ \bibnamefont
  {Po}}, \bibinfo {author} {\bibfnamefont {L.}~\bibnamefont {Fidkowski}},
  \bibinfo {author} {\bibfnamefont {T.}~\bibnamefont {Morimoto}}, \bibinfo
  {author} {\bibfnamefont {A.~C.}\ \bibnamefont {Potter}}, \ and\ \bibinfo
  {author} {\bibfnamefont {A.}~\bibnamefont {Vishwanath}},\ }\href {\doibase
  10.1103/PhysRevX.6.041070} {\bibfield  {journal} {\bibinfo  {journal} {Phys.
  Rev. X}\ }\textbf {\bibinfo {volume} {6}},\ \bibinfo {pages} {041070}
  (\bibinfo {year} {2016})}\BibitemShut {NoStop}%
\bibitem [{\citenamefont {Titum}\ \emph {et~al.}(2016)\citenamefont {Titum},
  \citenamefont {Berg}, \citenamefont {Rudner}, \citenamefont {Refael},\ and\
  \citenamefont {Lindner}}]{TitumFloquetAnderson}%
  \BibitemOpen
  \bibfield  {author} {\bibinfo {author} {\bibfnamefont {P.}~\bibnamefont
  {Titum}}, \bibinfo {author} {\bibfnamefont {E.}~\bibnamefont {Berg}},
  \bibinfo {author} {\bibfnamefont {M.~S.}\ \bibnamefont {Rudner}}, \bibinfo
  {author} {\bibfnamefont {G.}~\bibnamefont {Refael}}, \ and\ \bibinfo {author}
  {\bibfnamefont {N.~H.}\ \bibnamefont {Lindner}},\ }\href {\doibase
  10.1103/PhysRevX.6.021013} {\bibfield  {journal} {\bibinfo  {journal} {Phys.
  Rev. X}\ }\textbf {\bibinfo {volume} {6}},\ \bibinfo {pages} {021013}
  (\bibinfo {year} {2016})}\BibitemShut {NoStop}%
\bibitem [{\citenamefont {Nathan}\ \emph {et~al.}(2016)\citenamefont {Nathan},
  \citenamefont {Rudner}, \citenamefont {Lindner}, \citenamefont {Berg},\ and\
  \citenamefont {Refael}}]{nathan2016quantized}%
  \BibitemOpen
  \bibfield  {author} {\bibinfo {author} {\bibfnamefont {F.}~\bibnamefont
  {Nathan}}, \bibinfo {author} {\bibfnamefont {M.~S.}\ \bibnamefont {Rudner}},
  \bibinfo {author} {\bibfnamefont {N.~H.}\ \bibnamefont {Lindner}}, \bibinfo
  {author} {\bibfnamefont {E.}~\bibnamefont {Berg}}, \ and\ \bibinfo {author}
  {\bibfnamefont {G.}~\bibnamefont {Refael}},\ }\href@noop {} {\bibfield
  {journal} {\bibinfo  {journal} {arXiv preprint arXiv:1610.03590}\ } (\bibinfo
  {year} {2016})}\BibitemShut {NoStop}%
\bibitem [{\citenamefont {Fausti}\ \emph {et~al.}(2011)\citenamefont {Fausti},
  \citenamefont {Tobey}, \citenamefont {Dean}, \citenamefont {Kaiser},
  \citenamefont {Dienst}, \citenamefont {Hoffmann}, \citenamefont {Pyon},
  \citenamefont {Takayama}, \citenamefont {Takagi},\ and\ \citenamefont
  {Cavalleri}}]{fausti2011light}%
  \BibitemOpen
  \bibfield  {author} {\bibinfo {author} {\bibfnamefont {D.}~\bibnamefont
  {Fausti}}, \bibinfo {author} {\bibfnamefont {R.}~\bibnamefont {Tobey}},
  \bibinfo {author} {\bibfnamefont {N.}~\bibnamefont {Dean}}, \bibinfo {author}
  {\bibfnamefont {S.}~\bibnamefont {Kaiser}}, \bibinfo {author} {\bibfnamefont
  {A.}~\bibnamefont {Dienst}}, \bibinfo {author} {\bibfnamefont {M.~C.}\
  \bibnamefont {Hoffmann}}, \bibinfo {author} {\bibfnamefont {S.}~\bibnamefont
  {Pyon}}, \bibinfo {author} {\bibfnamefont {T.}~\bibnamefont {Takayama}},
  \bibinfo {author} {\bibfnamefont {H.}~\bibnamefont {Takagi}}, \ and\ \bibinfo
  {author} {\bibfnamefont {A.}~\bibnamefont {Cavalleri}},\ }\href@noop {}
  {\bibfield  {journal} {\bibinfo  {journal} {science}\ }\textbf {\bibinfo
  {volume} {331}},\ \bibinfo {pages} {189} (\bibinfo {year}
  {2011})}\BibitemShut {NoStop}%
\bibitem [{\citenamefont {Wang}\ \emph {et~al.}(2013)\citenamefont {Wang},
  \citenamefont {Steinberg}, \citenamefont {Jarillo-Herrero},\ and\
  \citenamefont {Gedik}}]{wang2013observation}%
  \BibitemOpen
  \bibfield  {author} {\bibinfo {author} {\bibfnamefont {Y.}~\bibnamefont
  {Wang}}, \bibinfo {author} {\bibfnamefont {H.}~\bibnamefont {Steinberg}},
  \bibinfo {author} {\bibfnamefont {P.}~\bibnamefont {Jarillo-Herrero}}, \ and\
  \bibinfo {author} {\bibfnamefont {N.}~\bibnamefont {Gedik}},\ }\href@noop {}
  {\bibfield  {journal} {\bibinfo  {journal} {Science}\ }\textbf {\bibinfo
  {volume} {342}},\ \bibinfo {pages} {453} (\bibinfo {year}
  {2013})}\BibitemShut {NoStop}%
\bibitem [{\citenamefont {Levonian}\ \emph {et~al.}(2016)\citenamefont
  {Levonian}, \citenamefont {Goldman}, \citenamefont {Singh}, \citenamefont
  {Markham}, \citenamefont {Twitchen},\ and\ \citenamefont
  {Lukin}}]{levonian2016probing}%
  \BibitemOpen
  \bibfield  {author} {\bibinfo {author} {\bibfnamefont {D.}~\bibnamefont
  {Levonian}}, \bibinfo {author} {\bibfnamefont {M.}~\bibnamefont {Goldman}},
  \bibinfo {author} {\bibfnamefont {S.}~\bibnamefont {Singh}}, \bibinfo
  {author} {\bibfnamefont {M.}~\bibnamefont {Markham}}, \bibinfo {author}
  {\bibfnamefont {D.}~\bibnamefont {Twitchen}}, \ and\ \bibinfo {author}
  {\bibfnamefont {M.}~\bibnamefont {Lukin}},\ }\href@noop {} {\bibfield
  {journal} {\bibinfo  {journal} {Bulletin of the American Physical Society}\ }
  (\bibinfo {year} {2016})}\BibitemShut {NoStop}%
\bibitem [{\citenamefont {Zhang}\ \emph {et~al.}(2017)\citenamefont {Zhang},
  \citenamefont {Hess}, \citenamefont {Kyprianidis}, \citenamefont {Becker},
  \citenamefont {Lee}, \citenamefont {Smith}, \citenamefont {Pagano},
  \citenamefont {Potirniche}, \citenamefont {Potter}, \citenamefont
  {Vishwanath} \emph {et~al.}}]{zhang2017observation}%
  \BibitemOpen
  \bibfield  {author} {\bibinfo {author} {\bibfnamefont {J.}~\bibnamefont
  {Zhang}}, \bibinfo {author} {\bibfnamefont {P.}~\bibnamefont {Hess}},
  \bibinfo {author} {\bibfnamefont {A.}~\bibnamefont {Kyprianidis}}, \bibinfo
  {author} {\bibfnamefont {P.}~\bibnamefont {Becker}}, \bibinfo {author}
  {\bibfnamefont {A.}~\bibnamefont {Lee}}, \bibinfo {author} {\bibfnamefont
  {J.}~\bibnamefont {Smith}}, \bibinfo {author} {\bibfnamefont
  {G.}~\bibnamefont {Pagano}}, \bibinfo {author} {\bibfnamefont {I.-D.}\
  \bibnamefont {Potirniche}}, \bibinfo {author} {\bibfnamefont
  {A.}~\bibnamefont {Potter}}, \bibinfo {author} {\bibfnamefont
  {A.}~\bibnamefont {Vishwanath}},  \emph {et~al.},\ }\href@noop {} {\bibfield
  {journal} {\bibinfo  {journal} {Nature}\ }\textbf {\bibinfo {volume} {543}},\
  \bibinfo {pages} {217} (\bibinfo {year} {2017})}\BibitemShut {NoStop}%
\bibitem [{\citenamefont {Choi}\ \emph {et~al.}(2017)\citenamefont {Choi},
  \citenamefont {Choi}, \citenamefont {Landig}, \citenamefont {Kucsko},
  \citenamefont {Zhou}, \citenamefont {Isoya}, \citenamefont {Jelezko},
  \citenamefont {Onoda}, \citenamefont {Sumiya}, \citenamefont {Khemani} \emph
  {et~al.}}]{choi2017observation}%
  \BibitemOpen
  \bibfield  {author} {\bibinfo {author} {\bibfnamefont {S.}~\bibnamefont
  {Choi}}, \bibinfo {author} {\bibfnamefont {J.}~\bibnamefont {Choi}}, \bibinfo
  {author} {\bibfnamefont {R.}~\bibnamefont {Landig}}, \bibinfo {author}
  {\bibfnamefont {G.}~\bibnamefont {Kucsko}}, \bibinfo {author} {\bibfnamefont
  {H.}~\bibnamefont {Zhou}}, \bibinfo {author} {\bibfnamefont {J.}~\bibnamefont
  {Isoya}}, \bibinfo {author} {\bibfnamefont {F.}~\bibnamefont {Jelezko}},
  \bibinfo {author} {\bibfnamefont {S.}~\bibnamefont {Onoda}}, \bibinfo
  {author} {\bibfnamefont {H.}~\bibnamefont {Sumiya}}, \bibinfo {author}
  {\bibfnamefont {V.}~\bibnamefont {Khemani}},  \emph {et~al.},\ }\href@noop {}
  {\bibfield  {journal} {\bibinfo  {journal} {Nature}\ }\textbf {\bibinfo
  {volume} {543}},\ \bibinfo {pages} {221} (\bibinfo {year}
  {2017})}\BibitemShut {NoStop}%
\bibitem [{\citenamefont {Bordia}\ \emph {et~al.}(2017)\citenamefont {Bordia},
  \citenamefont {Luschen}, \citenamefont {Schneider}, \citenamefont {Knap},\
  and\ \citenamefont {Bloch}}]{bordia2016periodically}%
  \BibitemOpen
  \bibfield  {author} {\bibinfo {author} {\bibfnamefont {P.}~\bibnamefont
  {Bordia}}, \bibinfo {author} {\bibfnamefont {H.}~\bibnamefont {Luschen}},
  \bibinfo {author} {\bibfnamefont {U.}~\bibnamefont {Schneider}}, \bibinfo
  {author} {\bibfnamefont {M.}~\bibnamefont {Knap}}, \ and\ \bibinfo {author}
  {\bibfnamefont {I.}~\bibnamefont {Bloch}},\ }\href
  {http://dx.doi.org/10.1038/nphys4020} {\bibfield  {journal} {\bibinfo
  {journal} {Nat Phys}\ }\textbf {\bibinfo {volume} {advance online
  publication}},\  (\bibinfo {year} {2017})}\BibitemShut {NoStop}%
\bibitem [{\citenamefont {Weidinger}\ and\ \citenamefont
  {Knap}(2017)}]{floquetprethermalregime}%
  \BibitemOpen
  \bibfield  {author} {\bibinfo {author} {\bibfnamefont {S.~A.}\ \bibnamefont
  {Weidinger}}\ and\ \bibinfo {author} {\bibfnamefont {M.}~\bibnamefont
  {Knap}},\ }\href {http://dx.doi.org/10.1038/srep45382} {\bibfield  {journal}
  {\bibinfo  {journal} {Scientific Reports}\ }\textbf {\bibinfo {volume} {7}},\
  \bibinfo {pages} {45382 EP } (\bibinfo {year} {2017})}\BibitemShut {NoStop}%
\bibitem [{\citenamefont {Houck}\ \emph {et~al.}(2012)\citenamefont {Houck},
  \citenamefont {T{\"u}reci},\ and\ \citenamefont {Koch}}]{houck2012chip}%
  \BibitemOpen
  \bibfield  {author} {\bibinfo {author} {\bibfnamefont {A.~A.}\ \bibnamefont
  {Houck}}, \bibinfo {author} {\bibfnamefont {H.~E.}\ \bibnamefont
  {T{\"u}reci}}, \ and\ \bibinfo {author} {\bibfnamefont {J.}~\bibnamefont
  {Koch}},\ }\href@noop {} {\bibfield  {journal} {\bibinfo  {journal} {Nature
  Physics}\ }\textbf {\bibinfo {volume} {8}},\ \bibinfo {pages} {292} (\bibinfo
  {year} {2012})}\BibitemShut {NoStop}%
\bibitem [{\citenamefont {Schecter}\ \emph {et~al.}(2012)\citenamefont
  {Schecter}, \citenamefont {Gangardt},\ and\ \citenamefont
  {Kamenev}}]{schecter2012dynamics}%
  \BibitemOpen
  \bibfield  {author} {\bibinfo {author} {\bibfnamefont {M.}~\bibnamefont
  {Schecter}}, \bibinfo {author} {\bibfnamefont {D.}~\bibnamefont {Gangardt}},
  \ and\ \bibinfo {author} {\bibfnamefont {A.}~\bibnamefont {Kamenev}},\
  }\href@noop {} {\bibfield  {journal} {\bibinfo  {journal} {Annals of
  Physics}\ }\textbf {\bibinfo {volume} {327}},\ \bibinfo {pages} {639}
  (\bibinfo {year} {2012})}\BibitemShut {NoStop}%
\bibitem [{\citenamefont {Meinert}\ \emph {et~al.}(2016)\citenamefont
  {Meinert}, \citenamefont {Knap}, \citenamefont {Kirilov}, \citenamefont
  {Jag-Lauber}, \citenamefont {Zvonarev}, \citenamefont {Demler},\ and\
  \citenamefont {N{\"a}gerl}}]{meinert2016bloch}%
  \BibitemOpen
  \bibfield  {author} {\bibinfo {author} {\bibfnamefont {F.}~\bibnamefont
  {Meinert}}, \bibinfo {author} {\bibfnamefont {M.}~\bibnamefont {Knap}},
  \bibinfo {author} {\bibfnamefont {E.}~\bibnamefont {Kirilov}}, \bibinfo
  {author} {\bibfnamefont {K.}~\bibnamefont {Jag-Lauber}}, \bibinfo {author}
  {\bibfnamefont {M.~B.}\ \bibnamefont {Zvonarev}}, \bibinfo {author}
  {\bibfnamefont {E.}~\bibnamefont {Demler}}, \ and\ \bibinfo {author}
  {\bibfnamefont {H.-C.}\ \bibnamefont {N{\"a}gerl}},\ }\href@noop {}
  {\bibfield  {journal} {\bibinfo  {journal} {arXiv preprint arXiv:1608.08200}\
  } (\bibinfo {year} {2016})}\BibitemShut {NoStop}%
\bibitem [{\citenamefont {Baumann}\ \emph {et~al.}(2010)\citenamefont
  {Baumann}, \citenamefont {Guerlin}, \citenamefont {Brennecke},\ and\
  \citenamefont {Esslinger}}]{baumann2010dicke}%
  \BibitemOpen
  \bibfield  {author} {\bibinfo {author} {\bibfnamefont {K.}~\bibnamefont
  {Baumann}}, \bibinfo {author} {\bibfnamefont {C.}~\bibnamefont {Guerlin}},
  \bibinfo {author} {\bibfnamefont {F.}~\bibnamefont {Brennecke}}, \ and\
  \bibinfo {author} {\bibfnamefont {T.}~\bibnamefont {Esslinger}},\ }\href@noop
  {} {\bibfield  {journal} {\bibinfo  {journal} {Nature}\ }\textbf {\bibinfo
  {volume} {464}},\ \bibinfo {pages} {1301} (\bibinfo {year}
  {2010})}\BibitemShut {NoStop}%
\bibitem [{\citenamefont {Gopalakrishnan}\ \emph {et~al.}(2016)\citenamefont
  {Gopalakrishnan}, \citenamefont {Knap},\ and\ \citenamefont
  {Demler}}]{heatingmbldrivenSG}%
  \BibitemOpen
  \bibfield  {author} {\bibinfo {author} {\bibfnamefont {S.}~\bibnamefont
  {Gopalakrishnan}}, \bibinfo {author} {\bibfnamefont {M.}~\bibnamefont
  {Knap}}, \ and\ \bibinfo {author} {\bibfnamefont {E.}~\bibnamefont
  {Demler}},\ }\href {\doibase 10.1103/PhysRevB.94.094201} {\bibfield
  {journal} {\bibinfo  {journal} {Phys. Rev. B}\ }\textbf {\bibinfo {volume}
  {94}},\ \bibinfo {pages} {094201} (\bibinfo {year} {2016})}\BibitemShut
  {NoStop}%
\bibitem [{\citenamefont {Ducatez}\ and\ \citenamefont
  {Huveneers}(2016)}]{ducatez2016anderson}%
  \BibitemOpen
  \bibfield  {author} {\bibinfo {author} {\bibfnamefont {R.}~\bibnamefont
  {Ducatez}}\ and\ \bibinfo {author} {\bibfnamefont {F.}~\bibnamefont
  {Huveneers}},\ }\href@noop {} {\bibfield  {journal} {\bibinfo  {journal}
  {arXiv preprint arXiv:1607.07353}\ } (\bibinfo {year} {2016})}\BibitemShut
  {NoStop}%
\bibitem [{\citenamefont {Shirley}(1965)}]{shirleyTLS}%
  \BibitemOpen
  \bibfield  {author} {\bibinfo {author} {\bibfnamefont {J.~H.}\ \bibnamefont
  {Shirley}},\ }\href {\doibase 10.1103/PhysRev.138.B979} {\bibfield  {journal}
  {\bibinfo  {journal} {Phys. Rev.}\ }\textbf {\bibinfo {volume} {138}},\
  \bibinfo {pages} {B979} (\bibinfo {year} {1965})}\BibitemShut {NoStop}%
\bibitem [{\citenamefont {Blanes}\ \emph {et~al.}(2009)\citenamefont {Blanes},
  \citenamefont {Casas}, \citenamefont {Oteo},\ and\ \citenamefont
  {Ros}}]{blanes2009magnus}%
  \BibitemOpen
  \bibfield  {author} {\bibinfo {author} {\bibfnamefont {S.}~\bibnamefont
  {Blanes}}, \bibinfo {author} {\bibfnamefont {F.}~\bibnamefont {Casas}},
  \bibinfo {author} {\bibfnamefont {J.}~\bibnamefont {Oteo}}, \ and\ \bibinfo
  {author} {\bibfnamefont {J.}~\bibnamefont {Ros}},\ }\href@noop {} {\bibfield
  {journal} {\bibinfo  {journal} {Physics Reports}\ }\textbf {\bibinfo {volume}
  {470}},\ \bibinfo {pages} {151} (\bibinfo {year} {2009})}\BibitemShut
  {NoStop}%
\bibitem [{\citenamefont {Anderson}(1958)}]{absenceofdiffusionanderson}%
  \BibitemOpen
  \bibfield  {author} {\bibinfo {author} {\bibfnamefont {P.~W.}\ \bibnamefont
  {Anderson}},\ }\href {\doibase 10.1103/PhysRev.109.1492} {\bibfield
  {journal} {\bibinfo  {journal} {Phys. Rev.}\ }\textbf {\bibinfo {volume}
  {109}},\ \bibinfo {pages} {1492} (\bibinfo {year} {1958})}\BibitemShut
  {NoStop}%
\bibitem [{\citenamefont {Abrahams}\ \emph {et~al.}(1979)\citenamefont
  {Abrahams}, \citenamefont {Anderson}, \citenamefont {Licciardello},\ and\
  \citenamefont {Ramakrishnan}}]{scalingtheoryandersonlocalization}%
  \BibitemOpen
  \bibfield  {author} {\bibinfo {author} {\bibfnamefont {E.}~\bibnamefont
  {Abrahams}}, \bibinfo {author} {\bibfnamefont {P.~W.}\ \bibnamefont
  {Anderson}}, \bibinfo {author} {\bibfnamefont {D.~C.}\ \bibnamefont
  {Licciardello}}, \ and\ \bibinfo {author} {\bibfnamefont {T.~V.}\
  \bibnamefont {Ramakrishnan}},\ }\href {\doibase 10.1103/PhysRevLett.42.673}
  {\bibfield  {journal} {\bibinfo  {journal} {Phys. Rev. Lett.}\ }\textbf
  {\bibinfo {volume} {42}},\ \bibinfo {pages} {673} (\bibinfo {year}
  {1979})}\BibitemShut {NoStop}%
\bibitem [{\citenamefont {Mott}(1968)}]{mottconductivity}%
  \BibitemOpen
  \bibfield  {author} {\bibinfo {author} {\bibfnamefont {N.~F.}\ \bibnamefont
  {Mott}},\ }\href {\doibase 10.1080/14786436808223200} {\bibfield  {journal}
  {\bibinfo  {journal} {Philosophical Magazine}\ }\textbf {\bibinfo {volume}
  {17}},\ \bibinfo {pages} {1259} (\bibinfo {year} {1968})},\ \Eprint
  {http://arxiv.org/abs/http://dx.doi.org/10.1080/14786436808223200}
  {http://dx.doi.org/10.1080/14786436808223200} \BibitemShut {NoStop}%
\bibitem [{\citenamefont {MacKinnon}\ and\ \citenamefont
  {Kramer}(1981)}]{mackinnonkramer}%
  \BibitemOpen
  \bibfield  {author} {\bibinfo {author} {\bibfnamefont {A.}~\bibnamefont
  {MacKinnon}}\ and\ \bibinfo {author} {\bibfnamefont {B.}~\bibnamefont
  {Kramer}},\ }\href {\doibase 10.1103/PhysRevLett.47.1546} {\bibfield
  {journal} {\bibinfo  {journal} {Phys. Rev. Lett.}\ }\textbf {\bibinfo
  {volume} {47}},\ \bibinfo {pages} {1546} (\bibinfo {year}
  {1981})}\BibitemShut {NoStop}%
\bibitem [{\citenamefont {Lee}\ and\ \citenamefont
  {Fisher}(1981)}]{leetwoDanderson}%
  \BibitemOpen
  \bibfield  {author} {\bibinfo {author} {\bibfnamefont {P.~A.}\ \bibnamefont
  {Lee}}\ and\ \bibinfo {author} {\bibfnamefont {D.~S.}\ \bibnamefont
  {Fisher}},\ }\href {\doibase 10.1103/PhysRevLett.47.882} {\bibfield
  {journal} {\bibinfo  {journal} {Phys. Rev. Lett.}\ }\textbf {\bibinfo
  {volume} {47}},\ \bibinfo {pages} {882} (\bibinfo {year} {1981})}\BibitemShut
  {NoStop}%
\bibitem [{\citenamefont {Agarwal}\ \emph {et~al.}(2015)\citenamefont
  {Agarwal}, \citenamefont {Gopalakrishnan}, \citenamefont {Knap},
  \citenamefont {M\"uller},\ and\ \citenamefont {Demler}}]{agarwalmblprl}%
  \BibitemOpen
  \bibfield  {author} {\bibinfo {author} {\bibfnamefont {K.}~\bibnamefont
  {Agarwal}}, \bibinfo {author} {\bibfnamefont {S.}~\bibnamefont
  {Gopalakrishnan}}, \bibinfo {author} {\bibfnamefont {M.}~\bibnamefont
  {Knap}}, \bibinfo {author} {\bibfnamefont {M.}~\bibnamefont {M\"uller}}, \
  and\ \bibinfo {author} {\bibfnamefont {E.}~\bibnamefont {Demler}},\ }\href
  {\doibase 10.1103/PhysRevLett.114.160401} {\bibfield  {journal} {\bibinfo
  {journal} {Phys. Rev. Lett.}\ }\textbf {\bibinfo {volume} {114}},\ \bibinfo
  {pages} {160401} (\bibinfo {year} {2015})}\BibitemShut {NoStop}%
\bibitem [{\citenamefont {Agarwal}\ \emph {et~al.}(2017)\citenamefont
  {Agarwal}, \citenamefont {Altman}, \citenamefont {Demler}, \citenamefont
  {Gopalakrishnan}, \citenamefont {Huse},\ and\ \citenamefont
  {Knap}}]{agarwal2017rare}%
  \BibitemOpen
  \bibfield  {author} {\bibinfo {author} {\bibfnamefont {K.}~\bibnamefont
  {Agarwal}}, \bibinfo {author} {\bibfnamefont {E.}~\bibnamefont {Altman}},
  \bibinfo {author} {\bibfnamefont {E.}~\bibnamefont {Demler}}, \bibinfo
  {author} {\bibfnamefont {S.}~\bibnamefont {Gopalakrishnan}}, \bibinfo
  {author} {\bibfnamefont {D.~A.}\ \bibnamefont {Huse}}, \ and\ \bibinfo
  {author} {\bibfnamefont {M.}~\bibnamefont {Knap}},\ }\href@noop {} {\bibfield
   {journal} {\bibinfo  {journal} {Annalen der Physik}\ } (\bibinfo {year}
  {2017})}\BibitemShut {NoStop}%
\bibitem [{\citenamefont {Shevchenko}\ \emph {et~al.}(2010)\citenamefont
  {Shevchenko}, \citenamefont {Ashhab},\ and\ \citenamefont
  {Nori}}]{shevchenko2010landau}%
  \BibitemOpen
  \bibfield  {author} {\bibinfo {author} {\bibfnamefont {S.}~\bibnamefont
  {Shevchenko}}, \bibinfo {author} {\bibfnamefont {S.}~\bibnamefont {Ashhab}},
  \ and\ \bibinfo {author} {\bibfnamefont {F.}~\bibnamefont {Nori}},\
  }\href@noop {} {\bibfield  {journal} {\bibinfo  {journal} {Physics Reports}\
  }\textbf {\bibinfo {volume} {492}},\ \bibinfo {pages} {1} (\bibinfo {year}
  {2010})}\BibitemShut {NoStop}%
\bibitem [{\citenamefont {Son}\ \emph {et~al.}(2009)\citenamefont {Son},
  \citenamefont {Han},\ and\ \citenamefont {Chu}}]{sonLZsolution}%
  \BibitemOpen
  \bibfield  {author} {\bibinfo {author} {\bibfnamefont {S.-K.}\ \bibnamefont
  {Son}}, \bibinfo {author} {\bibfnamefont {S.}~\bibnamefont {Han}}, \ and\
  \bibinfo {author} {\bibfnamefont {S.-I.}\ \bibnamefont {Chu}},\ }\href
  {\doibase 10.1103/PhysRevA.79.032301} {\bibfield  {journal} {\bibinfo
  {journal} {Phys. Rev. A}\ }\textbf {\bibinfo {volume} {79}},\ \bibinfo
  {pages} {032301} (\bibinfo {year} {2009})}\BibitemShut {NoStop}%
\bibitem [{\citenamefont {Lee}\ and\ \citenamefont
  {Ramakrishnan}(1985)}]{leeramakrishnan}%
  \BibitemOpen
  \bibfield  {author} {\bibinfo {author} {\bibfnamefont {P.~A.}\ \bibnamefont
  {Lee}}\ and\ \bibinfo {author} {\bibfnamefont {T.~V.}\ \bibnamefont
  {Ramakrishnan}},\ }\href {\doibase 10.1103/RevModPhys.57.287} {\bibfield
  {journal} {\bibinfo  {journal} {Rev. Mod. Phys.}\ }\textbf {\bibinfo {volume}
  {57}},\ \bibinfo {pages} {287} (\bibinfo {year} {1985})}\BibitemShut
  {NoStop}%
\bibitem [{\citenamefont {Hofstadter}(1976)}]{hofstadter}%
  \BibitemOpen
  \bibfield  {author} {\bibinfo {author} {\bibfnamefont {D.~R.}\ \bibnamefont
  {Hofstadter}},\ }\href {\doibase 10.1103/PhysRevB.14.2239} {\bibfield
  {journal} {\bibinfo  {journal} {Phys. Rev. B}\ }\textbf {\bibinfo {volume}
  {14}},\ \bibinfo {pages} {2239} (\bibinfo {year} {1976})}\BibitemShut
  {NoStop}%
\bibitem [{\citenamefont {Aubry}\ and\ \citenamefont
  {Andr{\'e}}(1980)}]{aubry1980analyticity}%
  \BibitemOpen
  \bibfield  {author} {\bibinfo {author} {\bibfnamefont {S.}~\bibnamefont
  {Aubry}}\ and\ \bibinfo {author} {\bibfnamefont {G.}~\bibnamefont
  {Andr{\'e}}},\ }\href@noop {} {\bibfield  {journal} {\bibinfo  {journal}
  {Ann. Israel Phys. Soc}\ }\textbf {\bibinfo {volume} {3}},\ \bibinfo {pages}
  {18} (\bibinfo {year} {1980})}\BibitemShut {NoStop}%
\bibitem [{\citenamefont {Ganeshan}\ \emph {et~al.}(2015)\citenamefont
  {Ganeshan}, \citenamefont {Pixley},\ and\ \citenamefont
  {Das~Sarma}}]{ganeshansingleparticlemobilityedge}%
  \BibitemOpen
  \bibfield  {author} {\bibinfo {author} {\bibfnamefont {S.}~\bibnamefont
  {Ganeshan}}, \bibinfo {author} {\bibfnamefont {J.~H.}\ \bibnamefont
  {Pixley}}, \ and\ \bibinfo {author} {\bibfnamefont {S.}~\bibnamefont
  {Das~Sarma}},\ }\href {\doibase 10.1103/PhysRevLett.114.146601} {\bibfield
  {journal} {\bibinfo  {journal} {Phys. Rev. Lett.}\ }\textbf {\bibinfo
  {volume} {114}},\ \bibinfo {pages} {146601} (\bibinfo {year}
  {2015})}\BibitemShut {NoStop}%
\end{thebibliography}

%

\end{document}